\documentclass[journal,letter]{IEEEtran}

\usepackage{graphicx}
\usepackage{amssymb}
\usepackage{siunitx}
\usepackage{amsmath}
\usepackage{algorithm}
\usepackage{algpseudocode}
\usepackage{cite}
\usepackage{stfloats}
\usepackage{subfigure}
\usepackage{epstopdf}
\usepackage{mdwlist}
\usepackage{threeparttable}
\usepackage[all,2cell]{xy} \UseAllTwocells
\usepackage{booktabs}
\usepackage{fancyvrb}
\usepackage{balance}
\usepackage{tensor}
\usepackage{leftidx}
\usepackage{mathabx}
\usepackage{color}
\usepackage{csquotes}
\usepackage{pdflscape}
\usepackage{afterpage}
\usepackage{array}
\usepackage{cases}
\usepackage{comment}

\newtheorem{lemma}{Lemma}

\newtheorem{remark}{Remark}

\AtBeginDocument{}

\newcommand{\arx}{a_\mathrm{rx}}
\newcommand{\Tx}{\mathrm{Tx}}
\newcommand{\Rx}{\mathrm{Rx}}
\newcommand{\Ix}{\mathrm{Ix}}

\newcommand{\CDF}[2]{F_{{#1}}\left({#2}\right)}

\newcommand{\Tb}{T_\mathrm{b}}

\newcommand{\Tr}{T_\mathrm{r}}
\newcommand{\dI}{d_\mathrm{I}}
\newcommand{\Pe}{P_\mathrm{b}}
\newcommand{\erfc}{\mathrm{erfc}}

\newcommand{\erf}{\mathrm{erf}}
\newcommand{\Yr}{Y_\mathrm{r}}
\newcommand{\Yt}{Y_\mathrm{T}}
\newcommand{\YI}{Y_\mathrm{I}}
\newcommand{\yt}{y_\mathrm{T}}
\newcommand{\yI}{y_\mathrm{I}}
\newcommand{\Xt}{X_\mathrm{T}}
\newcommand{\XI}{X_\mathrm{I}}
\newcommand{\xt}{x_\mathrm{T}}
\newcommand{\xI}{x_\mathrm{I}}
\newcommand{\pdi}{p_{d_\mathrm{I}}}
\newcommand{\PMF}[2]{P_{{#1}}\left({#2}\right)}

\newcommand{\argmin}{\mathop{\mathrm{arg\,min}}}
\newcommand{\argmax}{\mathop{\mathrm{arg\,max}}}

\newcounter{eqncnt1}
\newcounter{eqnback1}

\makeatletter
\def\endthebibliography{%
	\def\@noitemerr{\@latex@warning{Empty `thebibliography' environment}}%
	\endlist
}
\makeatother
\begin{document}
	
	\title{\LARGE
		Optimal Detection Interval for Absorbing Receivers in Molecular Communication Systems with Interference
	}
	
	\author{Trang~Ngoc~Cao,~\IEEEmembership{Student~Member,~IEEE,
		} Nikola~Zlatanov,~\IEEEmembership{Member,~IEEE,} Phee~Lep~Yeoh,~\IEEEmembership{Member,~IEEE,} and~Jamie~S.~Evans,~\IEEEmembership{Senior Member,~IEEE}%
		\thanks{This paper was presented in part at the
			IEEE International Conference on Communications 2018 \cite{Tra:18:icc}.}%
		\thanks{T. N. Cao and J. S. Evans are with the Department of Electrical and Electronic
			Engineering, University of Melbourne, Melbourne, VIC 3010, Australia
			(e-mail: ngocc@student.unimelb.edu.au; jse@unimelb.edu.au).}
		\thanks{N. Zlatanov is  with the Department of Electrical and Computer Systems
			Engineering, Monash University, Melbourne, VIC 3800, Australia (e-mail:
			nikola.zlatanov@monash.edu).}
		\thanks{P. L. Yeoh is with the School of Electrical and Information Engineering, University of Sydney, Sydney, NSW 2006, Australia
			(e-mail: phee.yeoh@sydney.edu.au).}
	}%
	
	
	\maketitle
	
	\begin{abstract}
		We consider a molecular communication system comprised of a transmitter, an absorbing receiver, and an  interference source. Assuming amplitude modulation, we  analyze the dependence of the bit error rate (BER) on the detection interval, which  is the time within one transmission symbol interval during which the receiver is active  to absorb and detect the number of information-carrying molecules. We then propose efficient algorithms to determine the optimal detection interval that minimizes the BER of the molecular communication system assuming no inter-symbol interference (ISI).   Simulation and numerical evaluations are  provided to highlight further insights into the optimal results.  For example, we demonstrate that 
		the optimal detection interval can be very small compared to the transmission symbol interval. Moreover, our numerical results show that significant BER improvements 	are achieved  by using the 	optimal detection interval for systems without and with ISI. 
		
	\end{abstract}
	

	\section{Introduction}
	
	Molecular communications (MC) is an exciting new paradigm that   overcomes  fundamental limits of size and operating environments in traditional radio frequency (RF)-based  communication systems. Molecular communications is well-suited to  challenging environments such as tunnels, pipelines, or salt water, where RF waves suffer extreme attenuation \cite{FYE:16:CSTO}. In addition, molecular communications is biocompatible and therefore can be used in human bodies for health monitoring, disease detection, or  drug delivery \cite{NMWVS:12:JN}.

	A promising platform for molecular communications is nano-machines, which will be able to perform more complex tasks if they can mutually communicate. Since each nano-machine can  perform simple operations, an essential requirement in  molecular communications  is simplicity. For example, only simple modulation techniques  can be used in molecular communications, such as amplitude modulation, where information is embedded into the number of released molecules at the  transmitter. In addition, only simple receivers can be employed. Two types of simple molecular receivers for amplitude modulation have been proposed in the literature so far; a passive and an absorbing receiver. A passive receiver is a receiver that observes and counts molecules in the receiving area at a specific sampling  instant without disrupting the movement of the molecules. An absorbing receiver is a receiver that absorbs and counts molecules reaching the receiver within a given detection  interval. We note that both passive and absorbing receivers can be realized by artificial cells or nano-machines \cite{XHC:16:mat, LS:10:US}.

	In general, the performance of a molecular communication system can be improved by adjusting the sampling  instants for passive receivers or the detection  interval  for absorbing receivers. This paper investigates the latter and  proposes algorithms that optimize the detection interval of an absorbing receiver in order to minimize the bit error rate (BER) when the molecular communication system is affected by an unintended transmitter from another communication link.
	 This is motivated by the fact that if a molecular communication system for nano-machines would be deployed in  a real environment, the communication session would experience interference from  various external  sources such as biochemical processes, leaking vesicles, or other unintended transmitters \cite{NCS:14:JSAC}. In particular,  sensor networks may have multiple communication links using the same type of molecules and the same designs of transmitters and receivers since the options of suitable molecular types and their corresponding transceivers' designs, e.g., suitable sensors, in a specific environment can be limited and a unified design is  convenient to expand the network. Hence, transmitters from these communication links result in external interference to each other, which has not been considered in the literature. Since nano-machine require simplicity, we need to find a simple solution to mitigate the impact of the interference. 
	 Related works on BER in molecular communication systems with multiple transmitters include papers such as \cite{NCS:14:JSAC} and \cite{JCL:14:ICASSP}. However, in these works, the detection interval is equal to the transmission symbol interval.

	For passive receivers, the optimal sampling instant   at which the receiver  observes  the largest number of molecules within one transmission symbol interval was derived in \cite{LCP:13:JSAC} and \cite{NCS:14:GLOBECOM}. In  \cite{NCS:14:JNB}, a passive receiver that observes multiple sampling instants during each transmission symbol interval was considered and maximum-likelihood detection was applied across all observation samples. Thereby, it was observed that the BER decreases when the number of  samples increases, which is intuitive since more information is received. Recently,  approximate closed-form expressions for the optimal number of samples and the optimal position of each sample within one transmission symbol interval  that minimize the BER were  analyzed in \cite{NKK:17:ICT}. 
	
	For absorbing receivers, most existing works \cite{DH:16:NB,HYC:15:MBSC,SML:15:TWC}  assume that the detection  interval is equal to  the transmission symbol interval. Exceptions are the works in  \cite{NKK:17:ICT,WGQ:14:ICT,MYCA:12:GLOBECOM,Tuna:18:CL},  which considered variable detection intervals. However, \cite{NKK:17:ICT,WGQ:14:ICT,MYCA:12:GLOBECOM,Tuna:18:CL} did not consider the impact of external interference from an unintended transmitter when optimizing the detection interval in terms of the system performance and only considered the internally generated inter-symbol interference (ISI) and constant-mean noise from the environment. The detection interval was optimized for minimizing the BER in  \cite{NKK:17:ICT,WGQ:14:ICT,Tuna:18:CL}, and for maximizing the capacity in \cite{MYCA:12:GLOBECOM} \footnote{The  period length $\tau$ in which
		molecules are absorbed by the receivers and removed from the environment, i.e., the period that is not the detection interval, was investigated and defined in \cite{WGQ:14:ICT} as the cleanse time.}. Moreover, \cite{NKK:17:ICT,WGQ:14:ICT,MYCA:12:GLOBECOM} determined the optimal detection time interval by exhaustive search, whereas  \cite{Tuna:18:CL} derives an approximately optimal detection time interval. Whereas, in this work, we consider both ISI and external interference from an unintended transmitter, whose \emph{conditional mean in each symbol interval varies} depending if the unintended transmitter transmits bits ``0" or ``1". Thereby, we propose  two efficient algorithms to optimize the detection interval  for minimizing the BER in a one dimensional (1D) MC system, which finds application in  long narrow tube environments, and a three dimensional (3D) MC system, which finds application in  free-space environments. We consider the most simple case, i.e., the 1D system, and the most general case, i.e., the 3D system, as  the two dimensional system can be straightforwardly analyzed by using the same framework.

	In this paper, we use the Binomial distribution to accurately describe the number of received molecules at the absorbing receiver  \cite{NCS:14:INB,FRMG:17:ISIT}. In addition, the Poisson and Gaussian distributions are also used since they provide an  approximation  of  the number of received molecules which is much  easier to  analyze \cite{NCS:14:GLOBECOM,  TU:16:WCOML,  NKK:17:ICT, HYC:15:MBSC,SML:15:TWC,NCS:14:JNB,NCS:14:JSAC,JCL:14:ICASSP},~\cite{Dam:17:NB}. However, note that the accuracy of the Poisson and Gaussian distributions does not always hold, as discussed  in \cite{NCS:14:INB} and \cite{YC:14:EL}. We investigate the molecular communication system both in a 1D space as in \cite{MAG:14:JSAC} and \cite{KEC:14:JN} as well as in a 3D  space as in \cite{NCS:14:GLOBECOM,NKK:17:ICT,NCS:14:JNB}. In addition, we investigate the interesting case, from a practical perspective, of an interference source with an unknown location in a 1D system, which has not been considered in the literature so far.  Our numerical results show that using the optimal detection interval, obtained by our proposed algorithms, leads to high performance in terms of BER.

	The main contributions of this work can be summarized as follows:
	\begin{itemize}
		\item We derive the BER of a MC system affected by external interference from  another communication link in 1D and 3D environments, when the system is impaired and is not impaired by ISI and when maximum likelihood (ML) detection is used. We consider three cases, i.e., when the Binomial, Poisson, and Gaussian distributions are used for the analysis, respectively. 
		\item We optimize the detection interval and show that the system performance in terms of BER is improved significantly by  choosing a suitable detection interval, for which we design a simple algorithm.
		\item We   optimize the detection interval and improve the BER even when the interference is at an unknown location in a 1D system.
	\end{itemize} 
	
	This paper expands its conference version \cite{Tra:18:icc} where 
	the analysis with approximations, i.e., when the Poisson and Gaussian distributions are used, and the ISI impact on the system are not included.
	
	The remainder of this paper is organized as follows. In Section~\ref{sec:2}, we introduce the system and channel models for  1D and  3D environments. In  Section~\ref{sec:3}, we construct an optimization problem of the optimal detection interval and derive the BER of the systems. Section~\ref{sec:4} proposes algorithms that optimize the detection interval in terms of BER. Section~\ref{sec:5} extends the investigation of the optimal detection interval to an interference source at an unknown location. Numerical results are provided in Section~\ref{sec:6}, and Section~\ref{sec:7} concludes the paper.
	
	\section{System and Channel Models}\label{sec:2}
	
	In the following, we present the system and channel models for our proposed molecular communication systems with interference.
	
	\subsection{System Model}
	
\begin{figure}
		\centering
		\includegraphics[width=0.48\textwidth]{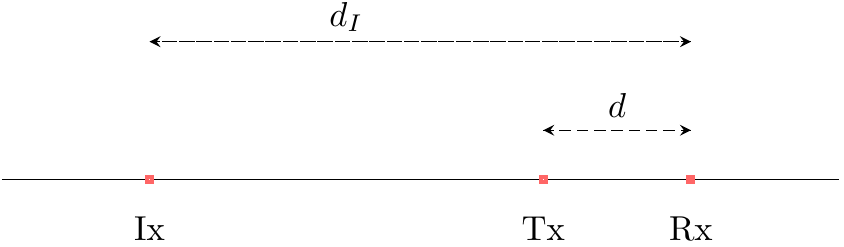}
		\caption{
			System model comprised of a  transmitter, $\mathrm{Tx}$, an interference source, $\mathrm{Ix}$, and a receiver, $\mathrm{Rx}$.
		}
		\label{fig:1}
	\end{figure}

	We consider a 1D unbounded MC system and a 3D unbounded MC system. The 1D system is comprised of  a point transmitter $\mathrm{Tx}$, a point absorbing receiver $\mathrm{Rx}$, and a point interference source $\mathrm{Ix}$.  The interference source $\mathrm{Ix}$ is assumed to be a transmitter from another communication link  that employs the same modulation and  molecule type as  $\mathrm{Tx}$.  $\mathrm{Rx}$ is assumed to be at  distances $d$  and  $\dI$ from $\mathrm{Tx}$ and $\mathrm{Ix}$, respectively, as shown in Fig.~\ref{fig:1}. The 3D system is comprised of  a spherical absorbing receiver $\mathrm{Rx}$ with radius $\arx$, a point transmitter $\mathrm{Tx}$ at a distance $d$ from the center of the receiver, and a point interference source $\mathrm{Ix}$ at a distance $d_\mathrm{I}$ from the center of the receiver. Note that the $\Tx$ and $\Ix$ do not need to be located on one side of the $\Rx$ in a 1D system or be aligned with $\Rx$ in a 3D system. The analysis in this paper applies to any relative positions of the transceivers that satisfies their respective distances.  We  assume that the  movement of the molecules in space follows  a Brownian motion \cite{AMG:17:MBSC}. We assume that both the intended and interfering transmitters,  $\mathrm{Tx}$ and $\mathrm{Ix}$, do not affect the diffusion of the molecules after they are released at the transmitters and that the receiver absorbs all molecules that reach it.
	
	We assume amplitude modulation, i.e., that  information bits  are modulated by the number of released molecules from  the transmitter. Let the number of released molecules at  $\mathrm{Tx}$ during the $j$-th transmission symbol interval be denoted by $\Xt^{(j)}$, where $\Xt^{(j)} \in \{N_0, N_1\}$, $N_1>N_0$. For brevity, we use $\Xt$ for arbitrary $j$, i.e., when there is no need to specify $j$. 
	When $\Xt=N_0$, then  bit \enquote{$0$} is assumed to be transmitted and when $\Xt=N_1$,   then  bit \enquote{$1$} is assumed to be transmitted. We consider $\Xt \in \{N_0, N_1\}$ instead of on-off keying to generalize the analysis so that it can be applied to higher-order modulation, e.g., $\Xt \in \{N_0, N_1, N_2, N_3\}$, in future work. This is also motivated by the fact that using a zero release rate of molecules  is not common in cell signaling in nature \cite{MPM:18:TCOM}.  
	We assume that the bits transmitted by  $\mathrm{Tx}$ are uncoded. As a result, the receiver is assumed to perform bit-by-bit detection of the received molecules. 
	For the interference transmitter, the number of molecules released by $\mathrm{Ix}$ during the transmission symbol interval $j$ is denoted by $\XI^{(j)}$, where  $\XI^{(j)}\in\left\{N_0,N_1\right\}$. Similar to $\Xt^{(j)}$, for brevity, we use $\XI$ for arbitrary $j$, i.e., when there is no need to specify $j$.
	We assume that the transmitted bits from $\Tx$ and $\Ix$ have equal-probabilities of occurrence defined, respectively, as
	\begin{align}\label{eq:1a}
	\PMF{X_\mathrm{T}}{x_\mathrm{T}=N_0}=\PMF{X_\mathrm{T}}{x_\mathrm{T}=N_1}=\frac{1}{2}
	\end{align} 
	and
	\begin{align}\label{eq:1b}
	\PMF{X_\mathrm{I}}{x_\mathrm{I}=N_0}=\PMF{X_\mathrm{I}}{x_\mathrm{I}=N_1}=\frac{1}{2}.
	\end{align}

	Let $\Tb$ denote the duration of the transmission symbol interval during which one information bit is transmitted at $\mathrm{Tx}$.  We assume instantaneous release and molecules  are released at the beginning of $\Tb$.  Let $\Tr$  denote the duration of the detection symbol interval during which  $\mathrm{Rx}$  absorbs and counts the number of absorbed  molecules in order to detect a transmitted information bit. We assume that  $\Tx$, $\Ix$, and $\Rx$ are synchronized, which can be done using already available techniques in the literature such as  the peak of the received molecular signal \cite{Muk:18:ants,Lin:18:WL}, the arrival time of the molecules \cite{Hsu:17:ICC}, the probability of molecules hitting a receiver \cite{Mit:19:icc},  a two-way message exchange between the two nanomachines \cite{Lin:16:Sen},  two types of molecules for synchronization and data transmission \cite{JAS:17:JNB, Muk:19:ICC}. Hence, we assume that   the transmission symbol intervals of $\Tx$ and $\Ix$ have the same duration and start at the same instant, which is also the starting instant of $\Tr$.
	
	\begin{remark}
		For a simple receiver without memory, the detection symbol intervals  cannot overlap with each other, i.e., each detection symbol interval  must start after the previous one has ended. In addition, $\Tr$ must be less or equal to $\Tb$, i.e., $\Tr \leq \Tb$ has to hold. Otherwise, if $\Tr > \Tb$, the detection symbol interval for the $j$-th transmitted bit ($j \gg 1$)  will start after a long period from the start of the  $j$-th bit  transmission symbol interval. In that case, the probability of receiving molecules belonging to the  $j$-th bit  approaches zero as $j$ increases, since most of those molecules would be absorbed in the previous detection symbol intervals.  Note that,  at time $t$,  where $\Tr<t<\Tb$, molecules should  still be absorbed at $\Rx$ in order to limit  ISI. However, we assume that these molecules are not included in the decision of the considered bit.
	\end{remark}

	\subsection{Channel Model}
	
	At the receiver, the information bits are detected based on the number of absorbed molecules during the detection symbol interval $\Tr$. Let $\Yt^{(j)}$ and $\YI^{(j)}$ denote the number of received molecules	at $\Rx$ during the interval $\Tr$ of the $j$-th bit which are released from $\Tx$ and $\Ix$ at the beginning of the $j$-th bit interval, respectively.
	Then, according to \cite{NCS:14:INB}, $\Yt^{(j)}$ and $\YI^{(j)}$ follow  Binomial distributions, i.e., $\Yt^{(j)} \sim \mathsf{Binom} \left(\Xt^{(j)}, p_d \right)$ and $\YI^{(j)} \sim \mathsf{Binom} \left(\XI^{(j)}, \pdi \right)$, respectively, where $\Xt^{(j)}$, $p_d$,  $\XI^{(j)}$, and $\pdi$ are parameters of the distributions. In particular, $p_d$ and  $\pdi$  are the probabilities that a  molecule released from $\Tx$ and $\Ix$ at the beginning of  $\Tb$ arrives at $\Rx$, placed at distance $d$  from $\Tx$  and $\dI$ from $\Ix$, within the interval $\Tr$, respectively.  Similar to $\Xt$ and $\XI$, for brevity, we use $\Yt$ and $\YI$ for arbitrary $j$, i.e., when there is no need to specify $j$.
	
	The probability mass function (PMF) of $Y_\mathrm{T}$ conditioned on $\Xt$ is given by
	\begin{align} \label{eq:16}
	\PMF{Y_\mathrm{T}|X_\mathrm{T}}{\left.y_\mathrm{T}\right|x_\mathrm{T}}=\binom{x_\mathrm{T}}{y_\mathrm{T}}p_d^{y_\mathrm{T}} (1-p_d)^{x_\mathrm{T}-y_\mathrm{T}}.
	\end{align}

	The PMF of $Y_\mathrm{I}$ conditioned on $\XI$ is then given by
	\begin{align}\label{eq:17}
	\PMF{Y_\mathrm{I}|X_\mathrm{I}}{\left.y_\mathrm{I}\right|x_\mathrm{I}}=\binom{x_\mathrm{I}}{y_\mathrm{I}}p_{d_\mathrm{I}}^{y_\mathrm{I}} (1-p_{d_\mathrm{I}})^{x_\mathrm{I}-y_\mathrm{I}}.
	\end{align}
	
	In a 1D unbounded environment,  
	$p_d$ and $\pdi$  are given, respectively, by \cite{FYE:16:CSTO}
	\begin{align} \label{eq:6}
	p_d=\erfc\left(\frac{d}{2\sqrt{D\Tr}}\right),
	\end{align}
	\begin{align} \label{eq:7}
	\pdi=\erfc\left(\frac{\dI}{2\sqrt{D\Tr}}\right),
	\end{align}
	where $\erfc(.)$ is the complementary error function and $D$ is the diffusion coefficient. 
	
	In a 3D unbounded environment, 
	$p_d$ and $p_{d_\mathrm{I}}$ are given respectively by \cite{FYE:16:CSTO, YHT:14:CL}
	\begin{align} \label{eq:27}
	p_d=\frac{r}{d}\erfc\left(\frac{d-r}{2\sqrt{D\Tr}}\right),
	\end{align}
	
	\begin{align} \label{eq:28}
	p_{d_\mathrm{I}}=\frac{r}{d_\mathrm{I}}\erfc\left(\frac{d_\mathrm{I}-r}{2\sqrt{D\Tr}}\right).
	\end{align}

	Since we consider $\Ix$ from an unintended transmitter, transmitting to a different absorbing receiver, the  absorbing receiver of the unintended transmitter can affect  the absorptions of the molecules transmitted from $\Tx$. Thus, the number of absorbed molecules at $\Rx$ may be reduced compared to the case when there is only one absorbing receiver. A few works have considered this effect \cite{Lu:16:NB,Ari:17:NB,Bao:19:WL}. In \cite{Lu:16:NB}, the interference receiver is assumed to be located at specific positions, i.e., aligned on a line or on a circle on the same plane with the transmitter and the target receiver. In \cite{Ari:17:NB}, the impact of the two receivers on each other was investigated by simulation. In \cite{Bao:19:WL}, a channel model was proposed based on a simulation fitting algorithm. However, in this work, we assume that this  effect is negligible. In fact, for the parameters chosen in this work, it is shown in Section~\ref{sec:6} that the impact is not significant. The results in the literature that investigate this effect can be  applied  in  our proposed framework by using the corresponding expressions of $p_d$ and $\pdi$ impacted by two receivers.

	In the following, we first formulate the general problem for optimizing the detection interval, $\Tr$, that minimizes the BER. We then assume the system is without ISI in order to derive the optimal detection and a tractable BER expression that can be used for optimizing $\Tr$. Next, we derive the optimal detection for the system with ISI and discuss the optimization of $\Tr$ for this system.

	\section{Problem Formulation and Detections}\label{sec:3}
	
	\subsection{Problem Formulation}
	
	The absorbing receiver detects the transmitted information based on the number of received molecules. Moreover, the numbers of  information and interference molecules received at $\mathrm{Rx}$, i.e., $\Yt$ and $\YI$, depend on $p_d$ and  $p_{d_\mathrm{I}}$, respectively, and thus depend on $\Tr$ due to \eqref{eq:27} and \eqref{eq:28}. Therefore, the BER of the system, denoted by $\Pe$, is a function of $\Tr$. Since $\Tr$ can be varied at the receiver, we can find the optimal detection interval $\Tr^\star$ that minimizes the BER. More precisely,  $\Tr^\star$ is found from the following optimization problem
	\begin{align} \label{eq:14}
	\Tr^\star=\underset{0\leq\Tr \leq \Tb}{\argmin}\Pe.
	\end{align}
	In order to solve the optimization problem in \eqref{eq:14}, we need to find the expression of the BER as a function of $\Tr$. 
	
	In order to focus on the effect of interference from  $\mathrm{Ix}$ and find a simple expression of the BER, we first assume that the ISI is negligible at the $\mathrm{Rx}$. This assumption becomes valid by setting the transmission symbol interval $\Tb$ to be long enough such that most of the molecules transmitted from  previous transmission symbol intervals  arrive at the $\mathrm{Rx}$, such as in \cite{KEC:14:JN} and \cite{TPY:15:MBSC}, or by using enzymes to react with the remaining molecules in the environment, such as in \cite{NCS:14:INB}. 
	
	In order for the detection process to be optimal, in terms of minimizing the BER, we consider ML detection at the receiver. We first consider a ML detection for the system assuming no ISI in order to solve the optimization problem \eqref{eq:14}. We then  consider a ML detection for the systems with ISI. 
	
	\subsection{Maximum Likelihood Detection without ISI}

	For the ML detection, the receiver decides whether $\Xt=N_0$ or $\Xt=N_1$ based on the following decision function
	\begin{align} \label{eq:8}
	\hat{X}_\mathrm{T}=\underset{\Xt\in\left\{N_0,N_1\right\}}{\argmax } \PMF{Y|\Xt}{\left.y\right|x_\mathrm{T}}, 
	\end{align}
	where $\hat{X}_\mathrm{T}$ is the detection of $\Xt$, $Y$ is the total number of  molecules received  at the receiver during the detection symbol interval $\Tr$  from both transmitters, given by
	\begin{align}\label{eq:3}
	Y=\Yt+\YI,
	\end{align}
	and  $\PMF{Y|\Xt}{\left.y\right|x_\mathrm{T}}$ is the conditional PMF of the total number of received molecules, $Y$, conditioned on the  number of transmitted molecules from  $\mathrm{Tx}$ being $\Xt$. Assuming no ISI, $\PMF{Y|\Xt}{\left.y\right|x_\mathrm{T}}$ can  be obtained as
	\begin{align} \label{eq:9}
	\PMF{Y|\Xt}{\left.y\right|x_\mathrm{T}}&=\PMF{Y|\Xt,X_\mathrm{I}}{\left.y\right|x_\mathrm{T},x_\mathrm{I}=N_0}\PMF{X_\mathrm{I}}{x_\mathrm{I}=N_0}\nonumber \\
	&+\PMF{Y|\Xt,X_\mathrm{I}}{\left.y\right|x_\mathrm{T},x_\mathrm{I}=N_1}\PMF{X_\mathrm{I}}{x_\mathrm{I}=N_1},
	\end{align}
	where $\PMF{Y|\Xt,X_\mathrm{I}}{\left.y\right|x_\mathrm{T},x_\mathrm{I}}$ is the conditional probability of receiving $Y$ molecules at the receiver when $X_\mathrm{T}$ and $X_\mathrm{I}$ molecules are released from the transmitters $\mathrm{Tx}$ and    $\mathrm{Ix}$, respectively, and $\PMF{X_\mathrm{I}}{x_\mathrm{I}}$ is the probability of releasing $X_\mathrm{I}$ molecules from $\mathrm{Ix}$.

	Let $\mathbb{Z}_0$ and $\mathbb{Z}_1$  be two sets comprised of numbers of received molecules, $Y$, for which the probability $\PMF{Y|\Xt}{\left.y\right|x_\mathrm{T}=N_0}$ is larger than the probability 
	$\PMF{Y|\Xt}{\left.y\right|x_\mathrm{T}=N_1}$ and vice versa, respectively. Then, \eqref{eq:8} is equivalent to the following 
	\begin{align} \label{eq:10}
	\hat{X}_\mathrm{T}=&\underset{\Xt\in\left\{N_0,N_1\right\}}{\argmax } \PMF{Y|\Xt}{\left.y\right|x_\mathrm{T}} \nonumber\\
	=&\begin{cases}
	N_0,  \text{if } \PMF{Y|\Xt}{\left.y\right|x_\mathrm{T}=N_0}>\PMF{Y|\Xt}{\left.y\right|x_\mathrm{T}=N_1}\\
	N_1,  \text{if } \PMF{Y|\Xt}{\left.y\right|x_\mathrm{T}=N_1	}\geq\PMF{Y|\Xt}{\left.y\right|x_\mathrm{T}=N_0}
	\end{cases} \nonumber \\
	=&
	\begin{cases}
	N_0,  \text{if } y \in \mathbb{Z}_0\\
	N_1,  \text{if } y \in \mathbb{Z}_1.
	\end{cases}
	\end{align}
	The sets $\mathbb{Z}_0$ and $\mathbb{Z}_1$ can be obtained by comparing $\PMF{Y|\Xt}{\left.y\right|x_\mathrm{T}=N_0}$ and 
	$\PMF{Y|\Xt}{\left.y\right|x_\mathrm{T}=N_1}$ for each $y$ in the interval $0 \leq y \leq 2N_1$. Note that the sets $\mathbb{Z}_0$ and $\mathbb{Z}_1$ can be calculated offline by the system designer and then stored at the receiver. For optimal detection, the receiver only needs to compare whether the received number of molecules, $Y$, belongs to the set  $\mathbb{Z}_0$ or the set $\mathbb{Z}_1$ and make a decision using  \eqref{eq:10}.
	Hence, the computational complexity of the proposed decision rule is low, which makes it suitable for a simple receiver. 
	
	Having defined the decision rule, given by \eqref{eq:10}, the BER can be obtained as
	\begin{align} \label{eq:11}
	\Pe&=\PMF{\hat{X}_\mathrm{T}|\Xt}{\left.\hat{x}_\mathrm{T}=N_0\right|x_\mathrm{T}=N_1} \PMF{\Xt}{x_\mathrm{T}=N_1}\nonumber \\ &+\PMF{\hat{X}_\mathrm{T}|\Xt}{\left.\hat{x}_\mathrm{T}=N_1\right|x_\mathrm{T}=N_0} \PMF{\Xt}{x_\mathrm{T}=N_0},
	\end{align}
	where $\PMF{\hat{X}_\mathrm{T}|\Xt}{\left.\hat{x}_\mathrm{T}\right|x_\mathrm{T}}$ is the PMF of detecting $\hat{X}_\mathrm{T}$ given that $X_\mathrm{T}$ was transmitted at $\mathrm{Tx}$, and
	$\PMF{X_\mathrm{T}}{x_\mathrm{T}}$ is the probability of releasing $X_\mathrm{T}$ molecules at $\mathrm{Tx}$.
	
	Now, to derive the BER as a function of $\Tr$ from \eqref{eq:11}, we first need to find $P_{\hat{X}_\mathrm{T}|\Xt}\left(\hat{x}_\mathrm{T}|x_\mathrm{T}\right)$. To this end, we use \eqref{eq:10}. Due to \eqref{eq:10}, we have
	\begin{align} \label{eq:12z}
	P_{\hat{X}_\mathrm{T}}\left(\hat{x}_\mathrm{T}=N_0\right)=\sum_{y\in \mathbb{Z}_0}P_Y\left(y\right),
	\end{align}
	and
	\begin{align} \label{eq:13z}
	P_{\hat{X}_\mathrm{T}}\left(\hat{x}_\mathrm{T}=N_1\right)=\sum_{y\in \mathbb{Z}_1}P_Y\left(y\right).
	\end{align}
	Thereby,
	\begin{align}\label{eq:12}
	\PMF{\hat{X}_\mathrm{T}|\Xt}{\left.\hat{x}_\mathrm{T}=N_0\right|x_\mathrm{T}=N_1}=\sum_{y\in \mathbb{Z}_0} \PMF{Y|\Xt}{\left.y\right|x_\mathrm{T}=N_1},
	\end{align}
	and
	\begin{align} \label{eq:13}
	\PMF{\hat{X}_\mathrm{T}|\Xt}{\left.\hat{x}_\mathrm{T}=N_1\right|x_\mathrm{T}=N_0}=\sum_{y\in \mathbb{Z}_1} \PMF{Y|\Xt}{\left.y\right|x_\mathrm{T}=N_0}.
	\end{align}
	
	Now, we need to obtain  $\PMF{Y|\Xt}{\left.y\right|x_\mathrm{T}}$ from \eqref{eq:9} and insert it into \eqref{eq:12} and \eqref{eq:13}. To this end, we first need to find $\PMF{Y|\Xt,\XI}{\left.y\right|x_\mathrm{T},x_\mathrm{I}}$. Since $\Yt$ and $\YI$ are independent, the PMF of $Y=\Yt+\YI$ can be found as a convolution of the PMFs of $\Yt$ given $X_\mathrm{T}$ and the PMF of $\YI$ given $X_\mathrm{I}$, as
	\begin{align} \label{eq:15a}
	\PMF{Y}{y}= \sum_{i=0}^{y}\PMF{\Yt}{i}\PMF{Y_\mathrm{I}}{y-i}. 
	\end{align}
	Conditioning both sides of \eqref{eq:15a} on $\Xt$ and $\XI$, we obtain 
	\begin{align} \label{eq:15b}
	\PMF{Y|\Xt,\XI}{\left.y\right|x_\mathrm{T},x_\mathrm{I}}=&\sum_{i=0}^{y}\PMF{\Yt|\Xt,\XI}{\left.i\right|x_\mathrm{T},x_\mathrm{I}}\\\nonumber
	&\times\PMF{Y_\mathrm{I}|\Xt,\XI}{\left.y-i\right|x_\mathrm{T},x_\mathrm{I}}.
	\end{align}
	Now, since $\Yt$ and $\YI$ are independent of $\XI$ and $\Xt$, respectively, \eqref{eq:15b}  can be written as 
	\begin{align} \label{eq:15}
	\PMF{Y|\Xt,X_\mathrm{I}}{\left.y\right|x_\mathrm{T},x_\mathrm{I}}= \sum_{i=0}^{y}\PMF{\Yt|\Xt}{\left.i\right|x_\mathrm{T}}\PMF{Y_\mathrm{I}|X_\mathrm{I}}{\left.y-i\right|x_\mathrm{I}}.
	\end{align}
	
	We now have all necessary expressions to write $\Pe$ in \eqref{eq:11} as a function of $\Tr$. To this end, we insert the PMF expressions in \eqref{eq:16} and \eqref{eq:17} into \eqref{eq:15}, then insert \eqref{eq:15} and \eqref{eq:1b} into \eqref{eq:9}, and obtain the conditional PMF $\PMF{Y|\Xt}{\left.y\right|x_\mathrm{T}}$. Finally, inserting $\PMF{Y|\Xt}{\left.y\right|x_\mathrm{T}}$ from \eqref{eq:9} into \eqref{eq:12} and  \eqref{eq:13} and then inserting them and \eqref{eq:1a} into \eqref{eq:11}, we derive the closed-form expression of the BER in \eqref{eq:peb}, given at the bottom of this page.
	We note that \eqref{eq:peb} is a general expression of the BER that holds for 1D and 3D environments by substituting the corresponding distributions for $p_d$ and $p_{d_\mathrm{I}}$ given in \eqref{eq:6}, \eqref{eq:7}, \eqref{eq:27}, and \eqref{eq:28}. 
	
		\setcounter{eqnback1}{\value{equation}} \setcounter{equation}{21}
		\begin{figure*} [!b]
			
			\hrulefill
				\begin{align} \label{eq:peb}
				&\Pe=\frac{1}{4}\times \nonumber\\ 
				&\left\{\sum_{y\in \mathbb{Z}_0}\sum_{i=0}^{y}\binom{N_1}{i}p_d^{i} (1-p_d)^{N_1-i}p_{d_\mathrm{I}}^{y-i}\left(\binom{N_0}{y-i} (1-p_{d_\mathrm{I}})^{N_0-\left(y-i\right)}+\binom{N_1}{y-i} (1-p_{d_\mathrm{I}})^{N_1-\left(y-i\right)}\right)\right.\nonumber\\ 
				&+\left.\sum_{y\in \mathbb{Z}_1}\sum_{i=0}^{y}\binom{N_0}{i}p_d^{i} (1-p_d)^{N_0-i}p_{d_\mathrm{I}}^{y-i}\left(\binom{N_0}{y-i} (1-p_{d_\mathrm{I}})^{N_0-\left(y-i\right)}+\binom{N_1}{y-i} (1-p_{d_\mathrm{I}})^{N_1-\left(y-i\right)}\right)\right\}
				\end{align}
				 \setcounter{equation}{24}
				\begin{align}\label{eq:9b}
				\PMF{Y|\mathbf{\Xt}^{(j)},\mathbf{\XI}^{(j)}}{y|\mathbf{\xt}^{(j)},\mathbf{\xI}^{(j)}}&=\PMF{\Yt^{(j)}|\Xt^{(j)}}{\yt^{(j)}|\xt^{(j)}}*\dots *\PMF{\Yt^{(j-L+1)}|\Xt^{(j-L+1)}}{\yt^{(j-L+1)}|\xt^{(j-L+1)}}\\\nonumber
				&*\PMF{\YI^{(j)}|\XI^{(j)}}{\yI^{(j)}|\xI^{(j)}}*\dots  *\PMF{\YI^{(j-L+1)}|\XI^{(j-L+1)}}{\yI^{(j-L+1)}|\xI^{(j-L+1)}}
				\end{align}
	\setcounter{eqncnt1}{\value{equation}}
	\setcounter{equation}{\value{eqnback1}}
\end{figure*}
	\subsection{Maximum Likelihood Detection with ISI}
	
	We now relax the assumption of negligible ISI in the previous subsection and consider ML detection for a channel with  memory $L$, i.e., the  molecules received at $\Rx$ during one bit interval are released from $\Tx$ and $\Ix$ during the current and $L-1$ previous bit intervals. Since we now consider a sequence of multiple bits, we use the superscript to denote the bit interval. The total number of received molecules during the detection interval of the $j$-th bit is then equal to
	\begin{align}\label{eq:8a}
	\setcounter{equation}{22}
	Y^{(j)}=\sum_{l=0}^{L-1}\Yt^{(j-l)}+\sum_{l=0}^{L-1}\YI^{(j-l)}.
	\end{align}
	$\PMF{Y|\Xt}{\left.y\right|x_\mathrm{T}}$ is now given by
	\begin{align} \label{eq:9a}
	\PMF{Y|\Xt^{(j)}}{\left.y\right|x_\mathrm{T}^{(j)}}&=\frac{1}{2^{2L-1}}\\\nonumber
	&\times\sum_{\mathbf{\Xt}^{(j)},\mathbf{\XI}^{(j)}}\PMF{Y|\mathbf{\Xt}^{(j)},\mathbf{\XI}^{(j)}}{y|\mathbf{\xt}^{(j)},\mathbf{\xI}^{(j)}},
	\end{align}
	where  $\mathbf{\Xt}^{(j)}=[\Xt^{(j-L+1)},\dots,\Xt^{(j)}]$, $\mathbf{\XI}^{(j)}=[\XI^{(j-L+1)},\dots,\XI^{(j)}]$, and the summation in \eqref{eq:9a} is over all possible values of $\mathbf{\Xt}^{(j)}$ and $\mathbf{\XI}^{(j)}$. $\PMF{Y|\mathbf{\Xt}^{(j)},\mathbf{\XI}^{(j)}}{y|\mathbf{\xt}^{(j)},\mathbf{\xI}^{(j)}}$ is given by \eqref{eq:9b} at the bottom of this page,
	where $*$ denotes convolution.
	Substituting \eqref{eq:9b} into \eqref{eq:9a}, we can obtain the ML detection from \eqref{eq:10} and the BER from \eqref{eq:11}, \eqref{eq:12}, \eqref{eq:13}, and \eqref{eq:9a}, respectively.

	The obtained BER expression is complicated for the ISI system and trying to optimize the BER in terms of $\Tr$ is computational expensive. Therefore,  optimizing $\Tr$ assuming negligible ISI  is more practical and can be considered as a suboptimal solution in the system with ISI. The performance of the systems without and with ISI using the optimal $\Tr$ obtained in the absence of ISI will be shown in Section~\ref{sec:6}.
	
	\sloppy In the above derivation of the closed-form expression of the BER for the non-ISI system, the probability $\PMF{Y|\Xt,\XI}{\left.y\right|x_\mathrm{T},x_\mathrm{I}}$ in \eqref{eq:15} can be derived  using the Binomial, Poisson, or Gaussian distribution. As explained in the introduction, the Binomial distribution describes the number of received molecules most accurately. The Poisson and Gaussian approximations are used in the literature due to their ease of analysis. In the next sections, we detail our proposed algorithms to obtain the optimal $\Tr$ according to \eqref{eq:14} and the corresponding BER for the three distributions. 
	
	\section{Optimal Receiving Interval in a System Affected by Interference at a Known Location} \label{sec:4}
	
	In this section, we propose algorithms to obtain the optimal detection interval, $\Tr^\star$, that minimizes the BER of the considered system model when the location of the interference source, $\mathrm{Ix}$, is known to the receiver, $\mathrm{Rx}$. We consider three cases, i.e., when the Binomial, Poisson, and Gaussian distributions are used for the analysis, respectively.

	\subsection{Optimizing $\Tr$ Using the Binomial Distribution}

	In order to develop an algorithm to optimize $\Tr$ in terms of $\Pe$, we need the observe the property of $\Pe$ as a function of $\Tr$.	
	From \eqref{eq:peb}, we can see that $\Pe$ is not a smooth function of $\Tr$ in general, since $ \mathbb{Z}_0$ and $ \mathbb{Z}_1$ change discretely as $\Tr$ changes. However, the following lemma will be useful for the algorithm development.
	\begin{lemma} \label{lem:1}
		There are intervals $T_\mathrm{r}^{(l)}\leq \Tr \leq T_\mathrm{r}^{(l+1)}$, for $l=1,2,\dots$, in which $\Pe$ is smooth with respect to $\Tr$.
	\end{lemma}
	    		\setcounter{eqnback1}{\value{equation}} \setcounter{equation}{26}
	    		\begin{figure*} [!b]
	    			
	    			\hrulefill
	    			\begin{align} \label{eq:pep}
	    			P_{\mathrm{b},\textrm{Poisson}}=&\frac{1}{4}\left\{\sum_{y\in \mathbb{Z}_0}\left(\frac{\left(N_1 p_d+N_0p_{d_\mathrm{I}}\right)^y e^{-\left(N_1 p_d+N_0p_{d_\mathrm{I}}\right)}}{y!}+\frac{\left(N_1 p_d+N_1p_{d_\mathrm{I}}\right)^y e^{-\left(N_1 p_d+N_1p_{d_\mathrm{I}}\right)}}{y!}\right)\right.\nonumber\\ 
	    			&+\left.\sum_{y\in \mathbb{Z}_1}\left(\frac{\left(N_0 p_d+N_0p_{d_\mathrm{I}}\right)^y e^{-\left(N_0 p_d+N_0p_{d_\mathrm{I}}\right)}}{y!}+\frac{\left(N_0 p_d+N_1p_{d_\mathrm{I}}\right)^y e^{-\left(N_0 p_d+N_1p_{d_\mathrm{I}}\right)}}{y!}\right)\right\},
	    			\end{align}
	    			\setcounter{eqncnt1}{\value{equation}}
	    			\setcounter{equation}{\value{eqnback1}}
	    		\end{figure*}    		
	\begin{proof}
		Please refer to Appendix~\ref{ap:0}.
	\end{proof}
	    		\begin{algorithm}[!t]
	    			\caption{Gradient projection method with steepest line search  for optimal detection interval using Binomial and Poisson distributions}\label{al1}
	    			\begin{algorithmic}[1] 
	    				\State $k\gets 0$, $t_{\mathrm{opt}}(0) \gets 0$, $\Tb$, $t_0$, $\epsilon \rightarrow 0$, $\alpha\in (0,1)$, $\beta \in (0,1)$, $c \in (0,1)$
	    				\While{$t_{\mathrm{opt}}(k) \leq \Tb$}
	    				\State $t_{\mathrm{min}}\gets t_0$
	    				\For {$y \in \{0,\dots,2N_1\}$}
	    				\If{$\PMF{\Yr|X}{\left.y\right|x=N_0} > \PMF{\Yr|X}{\left.y\right|x=N_1}$}
	    				\State $y \in \mathbb{Z}_0$
	    				\Else
	    				\State $y \in \mathbb{Z}_1$
	    				\EndIf
	    				\EndFor
	    				\State $t_{\mathrm{max}} : t\leq t_{\mathrm{max}}|\mathbb{Z}_0 \& \mathbb{Z}_1$ unchanged, {$t_{\mathrm{max}}$ found by binary search}
	    				\While{$\|t-t_1 \| \geq \epsilon$ }
	    				\State $t\gets t_1$
	    				\If{$t-\alpha \nabla \Pe(t) > t_{\mathrm{max}}$}
	    				\State $d \gets t_{\mathrm{max}}-t$
	    				\Else \If{$t-\alpha \nabla \Pe(t) < t_{\mathrm{min}}$}
	    				\State $d \gets t_{\mathrm{min}}-t$
	    				\Else
	    				\State $d \gets -\alpha \nabla \Pe(t)$
	    				\EndIf
	    				\EndIf
	    				\State $m \gets 0$
	    				\While{$\Pe(t) -\Pe(t+\beta^m d)<-c\beta^m\nabla \Pe(t)  d$}
	    				\State $m \gets m+1$
	    				\EndWhile
	    				\State $t_1\gets t+\beta^md$ 
	    				\EndWhile
	    				\State $t_{\mathrm{opt}}(k)\gets t_1$, $k\gets k+1$, $t_0\gets t_{\mathrm{max}}$
	    				\EndWhile
	    				\State $t^{\star} \gets \min(t_{\mathrm{opt}})$
	    				
	    			\end{algorithmic}
	    		\end{algorithm}	
    Given Lemma~\ref{lem:1}, we can find the optimal $\Tr$ in each of these intervals,  $T_\mathrm{r}^{(l)}\leq\Tr\leq T_\mathrm{r}^{(l+1)}$,  and obtain the corresponding minimal $\Pe$ for that interval and then compare the values of $\Pe$ from different intervals to find the global minimum. Algorithm~\ref{al1} outlines our proposed iterative algorithm for finding the optimal detection interval. In particular, we first specify the sets $ \mathbb{Z}_0$ and $ \mathbb{Z}_1$ for $T_\mathrm{r}^{(l)}$ (line 4-10) and  find $ T_\mathrm{r}^{(l+1)}$ such that $ \mathbb{Z}_0$ and $ \mathbb{Z}_1$ are fixed for $T_\mathrm{r}^{(l)}\leq\Tr\leq T_\mathrm{r}^{(l+1)}$ by binary search (line 11) \cite{CLRS:09:Book}. We then use gradient projection method (line 13-22) combined with steepest line  search satisfying Armijo rule (line 23-27)   \cite[Section~6.1]{Ber:15:Book} to find the optimal $\Tr$ in the interval $T_\mathrm{r}^{(l)}\leq\Tr\leq T_\mathrm{r}^{(l+1)}$. Finally, we find the global optimal  $\Tr$ by comparing optimal values of $\Tr$ in all intervals $T_\mathrm{r}^{(l)}\leq\Tr\leq T_\mathrm{r}^{(l+1)}$ (line 31).

	Note that  Algorithm~\ref{al1}  requires the gradient of the cost function, i.e., BER, and thus can only be used when the gradient of the BER is available. In other words,   Algorithm~\ref{al1} cannot be used to optimize the  detection interval of the systems with ISI due to the complicated BER expressions.

	\subsection{Approximation of the Optimal $\Tr$ Using the Poisson Distribution}
	
	When the number of released molecules is very large, i.e., $X_\mathrm{T} \gg 1$ and $X_\mathrm{I}\gg 1$ hold, the Binomial distributions of $\Yt$ and $\YI$ conditioned on $\Xt$ and $\XI$, respectively, can be approximated by Poisson distributions as $\Yt\sim \mathsf{Poiss}(\Xt p_d)$ and  $Y_\mathrm{I} \sim \mathsf{Poiss}(X_\mathrm{I}p_{d_\mathrm{I}})$  \cite{NCS:14:INB}. 
	
	Now,  due to the fact that the sum of two Poisson random variables also follows a Poisson distribution, we have $Y \sim \mathsf{Poiss}(\Xt p_d+X_\mathrm{I}p_{d_\mathrm{I}})$. Therefore,
	\setcounter{equation}{25}
	\begin{align} \label{eq:18} 
	\PMF{Y|\Xt,X_\mathrm{I}}{\left.y\right|x_\mathrm{T},x_\mathrm{I}}=\frac{\left(x_\mathrm{T} p_d+x_\mathrm{I}p_{d_\mathrm{I}}\right)^y e^{-\left(x_\mathrm{T} p_d+x_\mathrm{I}p_{d_\mathrm{I}}\right)}}{y!}.
	\end{align}
	Inserting \eqref{eq:18} and \eqref{eq:1b} into \eqref{eq:9}, then inserting \eqref{eq:9} into \eqref{eq:12},  \eqref{eq:13}, and then inserting them and \eqref{eq:1a} into \eqref{eq:11}, we obtain a closed-form expression for the BER in \eqref{eq:pep} given at the bottom of this page,
	where $p_d$ and $p_{d_\mathrm{I}}$ are given respectively by \eqref{eq:6} and \eqref{eq:7} for a 1D system, or \eqref{eq:27} and \eqref{eq:28} for a 3D system.

	Since the Poisson distribution is  discrete, we can use Algorithm~\ref{al1} to find the optimal $\Tr$.

	\subsection{Approximation of the Optimal $\Tr$ Using the Gaussian Distribution}
	
	Since $X_\mathrm{T} \gg 1$ and $X_\mathrm{I}\gg 1$ hold, the Binomial distributions of $\Yt$ and $\YI$ conditioned on $\Xt$ and $\XI$, respectively, can also be approximated  by Gaussian distributions as $\Yt \sim \mathcal{N}(\Xt p_d,\Xt p_d(1-p_d))$ and  $Y_\mathrm{I} \sim \mathcal{N}(X_\mathrm{I}p_{d_\mathrm{I}},X_\mathrm{I}p_{d_\mathrm{I}}(1-p_{d_\mathrm{I}}))$ \cite{NCS:14:INB}. In this case,  since the sum of two Gaussian random variables is also a Gaussian random variable, we have $Y \sim \mathcal{N}(\Xt p_d+X_\mathrm{I}p_{d_\mathrm{I}},\Xt p_d(1-p_d)+X_\mathrm{I}p_{d_\mathrm{I}}(1-p_{d_\mathrm{I}}))$ and
	\setcounter{equation}{27}
	\begin{align} \label{eq:19}
	&\PMF{Y|\Xt,X_\mathrm{I}}{\left.y\right|x_\mathrm{T},x_\mathrm{I}}\\\nonumber
	&\hspace{1cm}=\frac{1}{\sqrt{2\pi \left(x_\mathrm{T} p_d (1-p_d)+x_\mathrm{I}p_{d_\mathrm{I}}\left(1-p_{d_\mathrm{I}}\right)\right)}} \nonumber\\ 
	&\hspace{1cm}\times \exp{\left(-\frac{\left(y-\left(x_\mathrm{T} p_d+x_\mathrm{I}p_{d_\mathrm{I}}\right)\right)^2}{2\left(x_\mathrm{T} p_d (1-p_d)+x_\mathrm{I}p_{d_\mathrm{I}}\left(1-p_{d_\mathrm{I}}\right)\right)}\right)}.\nonumber
	\end{align}
	Then, $\PMF{Y|\Xt}{\left.y\right|x_\mathrm{T}}$ is found by inserting \eqref{eq:19} and \eqref{eq:1b} into \eqref{eq:9}. Note that, $\PMF{Y|\Xt,X_\mathrm{I}}{\left.y\right|x_\mathrm{T},x_\mathrm{I}}$ and $\PMF{Y|\Xt}{\left.y\right|x_\mathrm{T}}$ are now continuous functions with respect to $Y$ since $Y$ follows the Gaussian distribution. Therefore, $\mathbb{Z}_0$ and $\mathbb{Z}_1$, and the BER  now have to be derived differently than when $Y$ is  discrete.

	Since the set $\mathbb{Z}_k$, for $k \in\{0,1\}$ is now a continuous set, we can present $\mathbb{Z}_0$ and $\mathbb{Z}_1$ as a combination of the ranges $\left[\gamma_i ,\gamma_{i+1}\right]$, where $i$ is even for $k=0$ and $i$ is odd for $k=1$, and $\gamma_i$ and $\gamma_{i+1}$ are lower  and upper bounds of the range $i$. Then, from \eqref{eq:8}, we have $\PMF{Y|\Xt}{\left.y\right|x_\mathrm{T}=N_0}>\PMF{Y|\Xt}{\left.y\right|x_\mathrm{T}=N_1}$ when $y$ belongs to $\left[\gamma_i ,\gamma_{i+1}\right]$ and $i$ is even. Similarly, $\PMF{Y|\Xt}{\left.y\right|x_\mathrm{T}=N_1	}\geq\PMF{Y|\Xt}{\left.y\right|x_\mathrm{T}=N_0}$ holds when $y$ belongs to $\left[\gamma_i ,\gamma_{i+1}\right]$ and $i$ is odd. Therefore,
	$\gamma_i$, for $i=0,1,2,\dots$   are found by numerically solving  the following equation
	\begin{align} \label{eq:21}
	\PMF{Y|\Xt}{\left.y\right|x_\mathrm{T}=N_1}=\PMF{Y|\Xt}{\left.y\right|x_\mathrm{T}=N_0}.
	\end{align} 
	
	\setcounter{eqnback1}{\value{equation}} \setcounter{equation}{29}
	\begin{figure*} [!b]
		
		\hrulefill
		\begin{align} \label{eq:peg}
		&P_{\mathrm{b},\textrm{Gaussian}}=\frac{1}{8}\times \nonumber\\ 
		&\left(\sum_{i=0,2,\dots}\left(\erf\left(\frac{\gamma_{i+1}-\left(N_1p_d+N_0p_{d_\mathrm{I}}\right)}{N_1p_d(1-p_d)+N_0p_{d_\mathrm{I}}(1-p_{d_\mathrm{I}})}\right)
		-\erf\left(\frac{\gamma_i-\left(N_1p_d+N_0p_{d_\mathrm{I}}\right)}{N_1p_d(1-p_d)+N_0p_{d_\mathrm{I}}(1-p_{d_\mathrm{I}})}\right)\right.\right.\nonumber \\
		&+\left.\erf\left(\frac{\gamma_{i+1}-\left(N_1p_d+N_1p_{d_\mathrm{I}}\right)}{N_1p_d(1-p_d)+N_1p_{d_\mathrm{I}}(1-p_{d_\mathrm{I}})}\right)
		-\erf\left(\frac{\gamma_i-\left(N_1p_d+N_1p_{d_\mathrm{I}}\right)}{N_1p_d(1-p_d)+N_1p_{d_\mathrm{I}}(1-p_{d_\mathrm{I}})}\right)\right) \nonumber\\
		&+\sum_{i=1,3,\dots}\left(\erf\left(\frac{\gamma_{i+1}-\left(N_0p_d+N_0p_{d_\mathrm{I}}\right)}{N_0p_d(1-p_d)+N_0p_{d_\mathrm{I}}(1-p_{d_\mathrm{I}})}\right)
		-\erf\left(\frac{\gamma_i-\left(N_0p_d+N_0p_{d_\mathrm{I}}\right)}{N_0p_d(1-p_d)+N_0p_{d_\mathrm{I}}(1-p_{d_\mathrm{I}})}\right)\right.\nonumber \\
		&+\left.\left.\erf\left(\frac{\gamma_{i+1}-\left(N_0p_d+N_1p_{d_\mathrm{I}}\right)}{N_0p_d(1-p_d)+N_1p_{d_\mathrm{I}}(1-p_{d_\mathrm{I}})}\right)
		-\erf\left(\frac{\gamma_i-\left(N_0p_d+N_1p_{d_\mathrm{I}}\right)}{N_0p_d(1-p_d)+N_1p_{d_\mathrm{I}}(1-p_{d_\mathrm{I}})}\right)\right)\right),
		\end{align}
		
		\setcounter{eqncnt1}{\value{equation}}
		\setcounter{equation}{\value{eqnback1}}
	\end{figure*}

	The closed-form expression of the BER for this case is given in \eqref{eq:peg} at the bottom of this page, where $p_d$ and $p_{d_\mathrm{I}}$ are given respectively by \eqref{eq:6} and \eqref{eq:7} for a 1D system, or \eqref{eq:27} and \eqref{eq:28} for a 3D system (see Appendix~\ref{ap:1} for the detailed derivation). Since the bounds $\gamma_{i}$, for $i=0,1,2,\dots$, of set $\mathbb{Z}_0$ and set $\mathbb{Z}_1$ are found by numerically solving \eqref{eq:21}, i.e., there is no closed-form expression of $\gamma_{i}$,  deriving the derivative of the BER function does not lead to an insightful expression that can  be used in Algorithm~\ref{al1}. Therefore, we use implicit filtering \cite{NW:06:Book} to find the optimal detection interval, $\Tr^\star$, that minimizes the BER given in \eqref{eq:peg} as outlined in Algorithm~\ref{al2}. In particular, we use  implicit filtering \cite[Algorithm 9.6]{NW:06:Book} combined with projection (line 10 and 13) to ensure the new value of $\Tr$ is within the  range $[0,\Tb]$.

	\begin{remark}
		In general,  the Poisson approximation is more accurate than the Gaussian approximation when $p_d$ and $p_{d_\mathrm{I}}$ are   close to one or zero \cite{Dam:17:NB}, \cite{PP:02:Book}. In other cases, i.e., when  $p_d$ and $p_{d_\mathrm{I}}$ are not  close to one or zero, the Gaussian approximation is more accurate than the Poisson approximation. In practice, to keep the reliability of the system high, we must not design the system with $p_d$ close to zero, i.e., receiving very few information molecules, or $p_{d_\mathrm{I}}$ close to one, i.e., receiving too many interference molecules. Therefore, the Gaussian approximation may be more accurate in these designs despite the fact that Poisson approximation can capture the discreteness
		and non-negativity of the counting variable. 
	\end{remark}

	\begin{algorithm}[t!]
		\caption{Implicit filtering algorithm  for optimal detection interval using the Gaussian distribution}\label{al2}
		\begin{algorithmic}[1] 
			\State $\Tb$,  $a_{\mathrm{max}}$, $\epsilon \rightarrow 0$, $\tau \rightarrow 0$ $\alpha\in (0,1)$, $\beta \in (0,1)$, $c \in (0,1)$
			\While{$ \epsilon \geq \tau$}
			\State $increment \gets 0$
			\While{$increment = 0$ }
			\State $g\gets (\Pe(t+\epsilon)-\Pe(t+\epsilon))/(2\epsilon)$
			\If{$\|g\| \leq \epsilon$}
			\State $increment \gets 1$
			\Else 
			\State $m \gets 1$
			\State $d \gets \textbf{P}_{[0,\Tb]}(t-\rho^m g)$ (Projection of $t-\rho^m g$ on the value range of $\Tr$, $[0,\Tb]$)
			
			\While{$\Pe(d)>\Pe(t)-\alpha\beta^m g^2$ and $m<a_{\mathrm{max}}$}
			\State {$m \gets m+1$}
			\State $d \gets \textbf{P}_{[0,\Tb]}(t-\beta^m g)$ (Projection of $t-\beta^m g$ on the value range of $\Tr$, $[0,\Tb]$)
			\EndWhile
			\If{$m=a_{\mathrm{max}}$}
			\State $increment \gets 1$
			\Else
			\State $t=d$
			\EndIf
			\EndIf
			
			\EndWhile
			\State $\epsilon \gets \epsilon c$
			
			\EndWhile

		\end{algorithmic}
	\end{algorithm}

	\begin{remark}
		The optimal $\Tr$ given by Algorithm~\ref{al1} is a global optimum. Since the Binomial distribution is approximated by the Poisson and the Gaussian distributions, the three distributions result in similar behaviors of the BER (shown in the  numerical section). Therefore, we  give  proof for the  global optimum of $\Tr$ only for the case of  the Poisson distribution (See Appendix~\ref{ap:5}). 
	\end{remark}
	
	\begin{remark}
		In order to optimize the detection interval, we proposed  suitable algorithms according to the properties of the optimization problems. In particular, Algorithm~\ref{al1} and \ref{al2} handle the lack of the function smoothness and of the function derivative, respectively. The optimization process using these algorithms can be done offline and the result can then be used to set the optimal duration of the detection interval  at the receiver. Hence, there is no complex calculation required in the MC systems yet  system performance is improved by the proposed optimal design.
	\end{remark}

	\section{Optimal Receiving Interval in a System Affected by Interference at an Unknown Location} \label{sec:5}
	
	In this section, we generalize the investigation of the 1D system and consider that the exact location of the interference source $\mathrm{Ix}$ is unknown to the receiver $\mathrm{Rx}$. Instead, the receiver  has only statistical knowledge of the location. 
	
	We assume that the interference source is randomly located between distances $a$ and $b$ from the receiver according to the uniform distribution. Thereby, the distance $d_\mathrm{I}$ from the receiver to the interference source, $\mathrm{Ix}$, is now a random variable following the uniform distribution, i.e., $d_\mathrm{I} \sim \mathcal{U}(a,b)$. Since the receiver does not know $d_\mathrm{I}$, the detection process is optimal when the receiver uses maximum likelihood of the expectation of the PMF of the number of received molecules, as follows
	\setcounter{equation}{30}
	\begin{align} \label{eq:30}
	\hat{X}&=\underset{\Xt\in\left\{N_0,N_1\right\}}{\argmax } \mathbb{E}_{d_\mathrm{I}}\left[\PMF{Y|\Xt}{\left.y\right|x_\mathrm{T}}\right]\nonumber \\
	&=\underset{\Xt\in\left\{N_0,N_1\right\}}{\argmax } \int_a^b\frac{1}{b-a}\PMF{Y|\Xt}{\left.y\right|x_\mathrm{T}}dd_\mathrm{I}, 
	\end{align}
	where $\PMF{Y|\Xt}{\left.y\right|x_\mathrm{T}}$ is given as in Section~\ref{sec:4} for each corresponding distribution  and $\mathbb{E}\left[.\right]$ denotes the expectation.
	
	For the detection rule in this case, we redefine $\mathbb{Z}_0$ and $\mathbb{Z}_1$  as the sets of numbers  of received molecules for which $\mathbb{E}_{d_\mathrm{I}}\left[\PMF{Y|\Xt}{\left.y\right|x_\mathrm{T}=N_0}\right]$ is larger than $\mathbb{E}_{d_\mathrm{I}}\left[\PMF{Y|\Xt}{\left.y\right|x_\mathrm{T}=N_1}\right]$ and vice versa, respectively, when $0\leq y\leq 2N_1$. For both the Binomial and Poisson distributions, $\mathbb{Z}_0$ and $\mathbb{Z}_1$ can be found by comparing $\mathbb{E}_{d_\mathrm{I}}\left[\PMF{Y|\Xt}{\left.y\right|x_\mathrm{T}=N_0}\right]$ and  $\mathbb{E}_{d_\mathrm{I}}\left[\PMF{Y|\Xt}{\left.y\right|x_\mathrm{T}=N_1}\right]$. On the other hand, for Gaussian distribution, $\mathbb{Z}_0$ and $\mathbb{Z}_1$ can be found by numerically solving the following equation
	\begin{align}
	\int_a^b\PMF{Y|\Xt}{\left.y\right|x_\mathrm{T}=N_0}dd_\mathrm{I}=\int_a^b\PMF{Y|\Xt}{\left.y\right|x_\mathrm{T}=N_1}dd_\mathrm{I}.
	\end{align} 
	Furthermore, from \eqref{eq:30}, we have
	\begin{align} \label{eq:13c}
	&\PMF{\hat{X}_\mathrm{T}|\Xt}{\left.\hat{x}_\mathrm{T}=N_0\right|x_\mathrm{T}=N_1}\\\nonumber&\hspace{1cm}=\sum_{y\in \mathbb{Z}_0} \int_a^b\frac{1}{b-a}\PMF{Y|\Xt}{\left.y\right|x_\mathrm{T}=N_1}dd_\mathrm{I}
	\end{align}
	and
	\begin{align} \label{eq:13d}
	&\PMF{\hat{X}_\mathrm{T}|\Xt}{\left.\hat{x}_\mathrm{T}=N_1\right|x_\mathrm{T}=N_0}\\\nonumber&\hspace{1cm}=\sum_{y\in \mathbb{Z}_1} \int_a^b\frac{1}{b-a}\PMF{Y|\Xt}{\left.y\right|x_\mathrm{T}=N_0}dd_\mathrm{I}.
	\end{align}
	Therefore, using  similar derivation as in Section~\ref{sec:4} with $P_{\hat{X}_\mathrm{T}|\Xt}$  given by \eqref{eq:13c} and \eqref{eq:13d},
	we can obtain the BER $\Pe'$ of the system affected by interference at an unknown location as follows
	\begin{align} \label{eq:pe}
	\Pe'=\int_a^b\frac{1}{b-a}\Pe dd_\mathrm{I}.
	\end{align}
	Note that \eqref{eq:pe} holds for the corresponding BER for Binomial, Poisson, and Gaussian distributions.

	We can use the algorithms developed in Section~\ref{sec:4} to find the optimal detection interval when Binomial, Poisson, and Gaussian distributions are used, respectively. 
	
	Note that the results in this  section can be extended to the 3D case. In that case, the derivation in \eqref{eq:30}-\eqref{eq:pe} needs to evaluate the integration over the 3D space that the interference is located in, instead of the 1D integral.

	\section{Numerical Results} \label{sec:6}
	
	\begin{table}[t!]
		\caption{Parameters of the systems used for numerical results}
		\centering
		\label{table:1}
		\begin{tabular}{c|c||c|c}
			\toprule
			Parameter  & Value &Parameter    & Value \\[-0.05cm]
			\midrule
			$D$ [$\mathrm{m^2/s}$]	&$10^{-9}$&$r$ [$\mathrm{m}$]        &$1\times 10^{-6}$\\
			$d$ [$\mathrm{m}$]   &$1.5\times 10^{-5}$&    		$d_\mathrm{I}$ [$\mathrm{m}$]  &$6\times 10^{-5}$\\
			$\Tb$[\si{\second}](1D)& $7.12$ & $\Tb$[\si{\second}](3D)& $6.21$\\
			$N_0=M_0 (1D)$  &$20$ & $N_1=M_1(1D)$ & $40$\\
			$N_0=M_0 (3D)$  &$1000$&$N_1=M_1(3D)$ & $2000$\\
			\bottomrule
		\end{tabular}
	\end{table}
	
\begin{figure}
		\centering
		\includegraphics[width=0.450\textwidth]{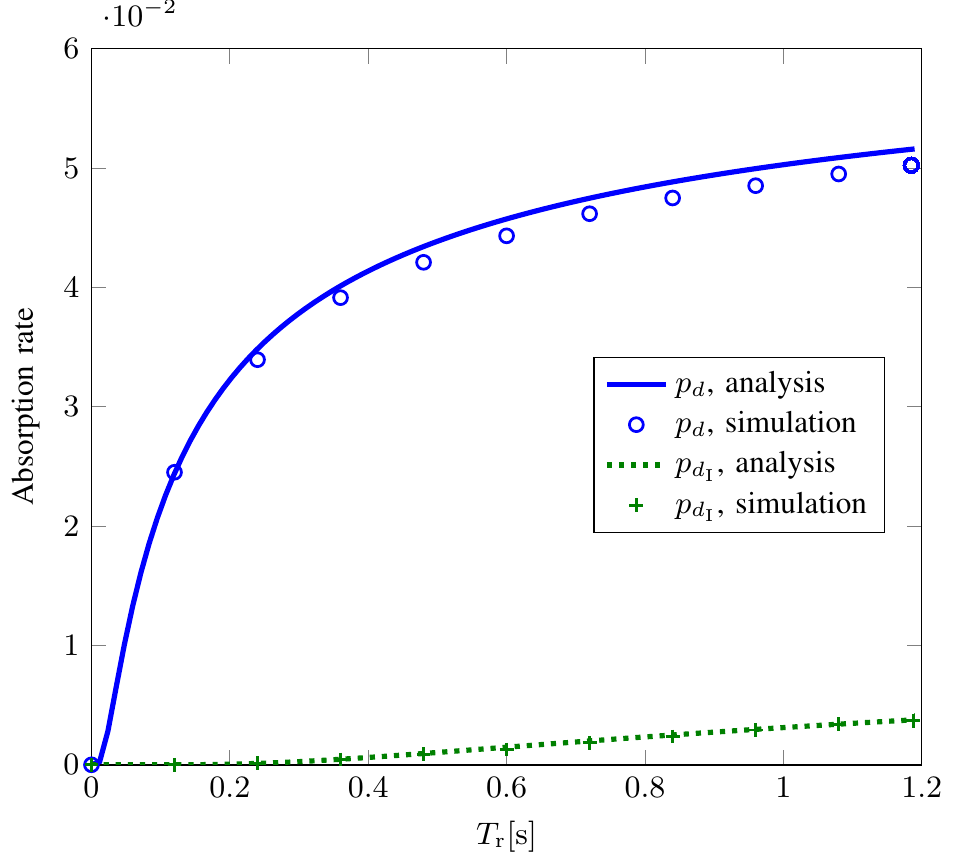}
		\caption{
			Absorption rates, $p_d$ and $\pdi$, as a function of $\Tr$ when there are two pairs of transceivers in a 3D MC system.
		}
		\label{fig:3}
	\end{figure}

	In this section, we  illustrate the dependence of the BER  on the detection  interval $\Tr$ and show the impacts of optimizing $\Tr$ on the BER. Unless otherwise stated, we use the default values of the parameters given in Table~\ref{table:1}. In an unbounded 3D environment, larger amounts of molecules are needed since the molecules diffuse in all dimensions and only a small portion of them can reach the receiver. For the system parameters in Table~\ref{table:1}, to ensure that the ISI caused by the $\Tx$ is small, $\Tb$ is chosen such that the ratio between $p_d$ with $\Tr=\Tb$ to $p_d$ with $\Tr\rightarrow \infty$ is equal to $90\%$. Eliminating interference caused by the $\Ix$ will be taken into account in the design of the optimal detection interval. For smaller $d$, the ISI caused by the $\Tx$ will be higher compared to the default value of $d$ in Table~\ref{table:1}. We will highlight in our results that the design of the optimal detection interval to eliminate interference from $\Ix$ is still valid for the performance improvement of the system with ISI. Unless otherwise stated,  the  value of $\Tb$ is fixed in order to investigate the impact of $\Tr$.  The detection interval is optimized with the assumption of no ISI. However, the performance of systems with ISI is also shown numerically. For Fig.~\ref{fig:3}, we adopt the particle-based simulation of Brownian motion, where the molecules take a random step in space for every discrete time step of length $\Delta t=10^{-5} \si{\second}$. The length of  each step in each spatial dimension  is modeled as a Gaussian random variable with zero mean and standard deviation  $\sqrt{2D\Delta t}$.
	In the other simulation, we  adopt Monte-Carlo simulation   by averaging the BER over  $10^5$ transmissions. In particular, we generate  released molecules according to the modulation rule, counting the number of  molecules absorbed during the detection interval. Then, the decoded bit is decided by comparing whether the number of received molecules $Y$   belongs to the set $\mathbb{Z}_0$ or   $\mathbb{Z}_1$, as in \eqref{eq:10}.


	In Fig.~\ref{fig:3}, we consider a 3D  MC system with two pairs of transceivers to present the case when the impact of an absorbing receiver on the other is not significant and thus verify our assumption.  In Fig.~\ref{fig:3}, we plot $p_d$ and $\pdi$ as  functions of the detection interval $\Tr$ with analytical expressions given in \eqref{eq:27} and \eqref{eq:28}, respectively. We observe that $p_d$ given by  \eqref{eq:27} matches the simulation points. We also observe an exact match for $\pdi$.

	\begin{figure}
		\centering
		\includegraphics[width=0.450\textwidth]{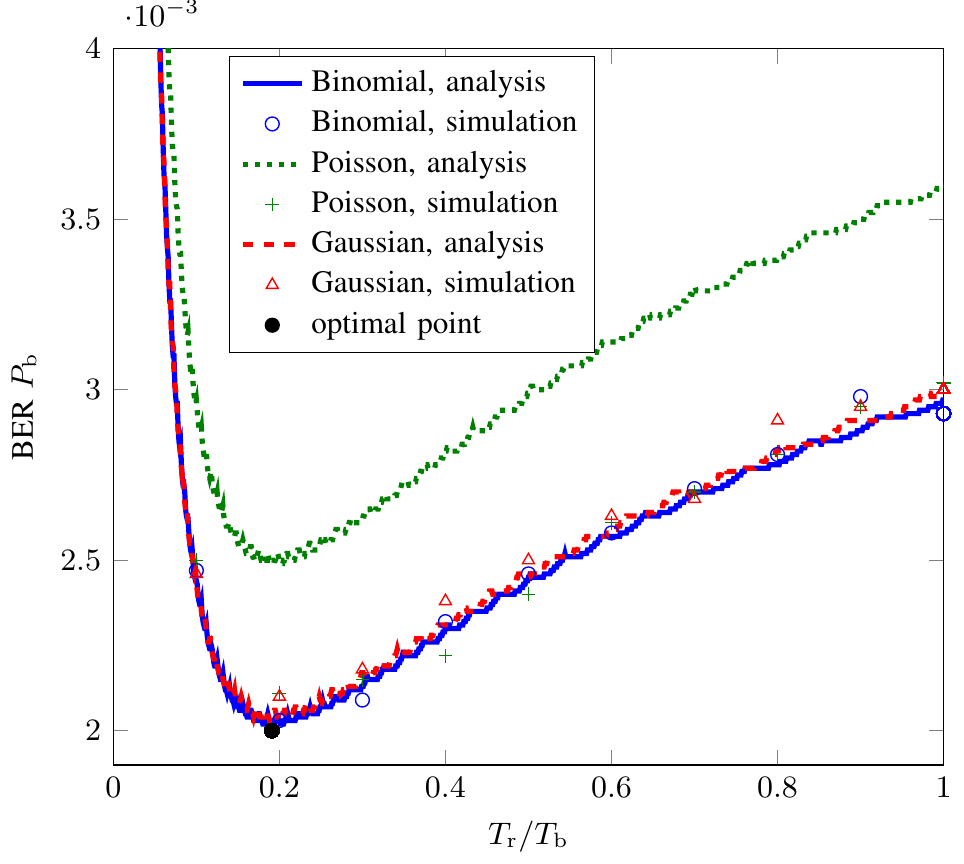}
		\caption{
			BER  as a function of $\Tr/\Tb$ in a 3D system affected by interference when the number of received molecules is described by a Binomial distribution and approximated by Poisson and Gaussian distributions.
		}
		\label{fig:2}
	\end{figure}

	Fig.~\ref{fig:2} shows the BER of a 3D system as a function of the ratio of the detection interval, $\Tr$, to the transmission symbol interval, $\Tb$, when the number of received molecules is described by the Binomial distribution and when it is approximated by Poisson and Gaussian distributions. As can be seen from Fig.~\ref{fig:2}, the BER in the system affected by external interference does not  decrease monotonically when $\Tr$ increases. Thereby the optimal detection interval, $\Tr^\star$, that minimizes the BER is usually not  equal to the transmission symbol interval, $\Tb$. In fact, when $\Tr$ increases and $\Tb$ is constant, the BER decreases to a minimum value and then increases. 
	The minimum value of the BER matches with the BER of the optimal $\Tr$ found by Algorithm~\ref{al1},  i.e., the black dot in Fig.~\ref{fig:2}.
	The dependence of the BER on $\Tr$ can be explained as follows. When $\Tr=0$, $\Pe=0.5$ since there are no received molecules at the $\mathrm{Rx}$. As $\Tr$ increases, more molecules from $\Tx$ are received at the $\mathrm{Rx}$ and thus the BER decreases from the maximum of $0.5$ when $\Tr/\Tb=0$ and reaches a minimum value of $\SI{2e-3}{}$ when $\Tr/\Tb = 0.2$.  When $\Tr$ increases even more, more transmitted molecules from $\mathrm{Tx}$ and $\mathrm{Ix}$ are received and the impact of molecules from $\Ix$ becomes more significant. Therefore, the BER increases. Moreover, we observe a mismatch between the analysis results with Poisson distribution and other results since Poisson is only an accurate approximation when $p_d$ and $\pdi$ are close to $0$ or $1$ \cite{PP:02:Book}. The simulation results are in agreement since we generate the number of received molecules following the true Binomial distribution and only use the approximated distribution for  detection designs. However, the analytical results for the Poisson distribution display similar characteristics with other results as a function of $\Tr/\Tb$, i.e., has the same minimum point. 
	
	\begin{figure}
		\centering
		\includegraphics[width=0.45\textwidth]{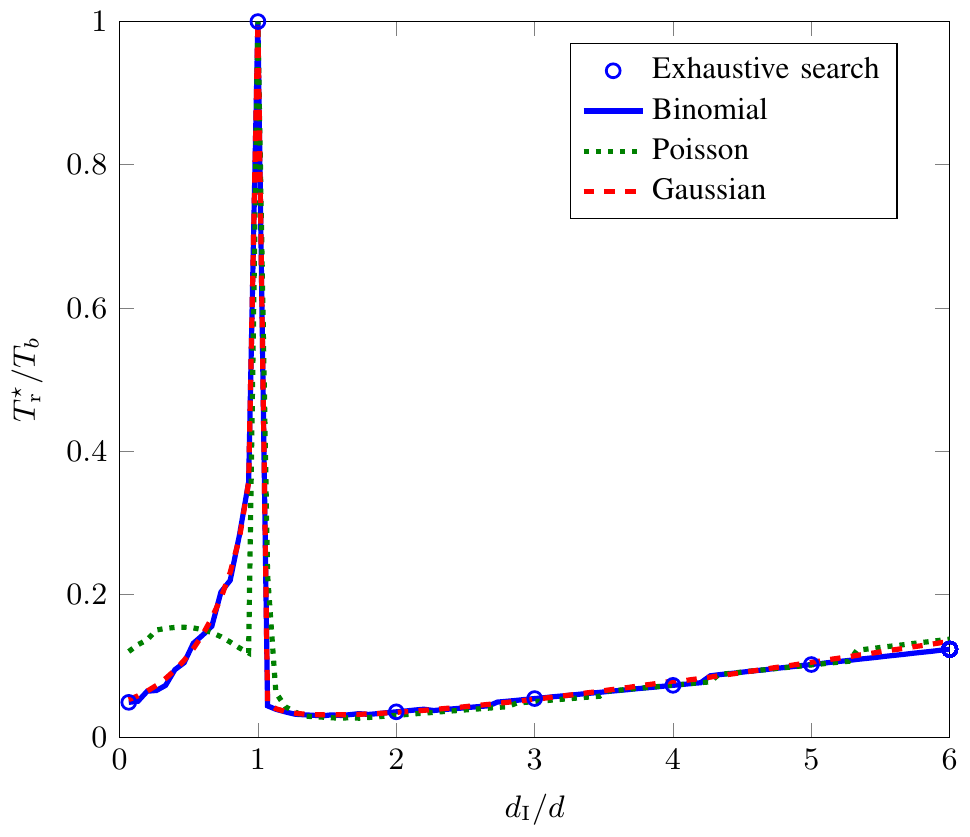}
		\caption{
			The ratio of the optimal detection  interval, $\Tr^\star$, to the transmission symbol interval, $\Tb$, as a function of the ratio of $d_\mathrm{I}$ to $d$ in a 1D system when using the Binomial distribution, and Poisson and Gaussian approximations.
		}
		\label{fig:4}
	\end{figure}

	In Fig.~\ref{fig:4}, the ratio of the optimal detection  interval, $\Tr^\star$, to a fixed transmission symbol interval, $\Tb$, is shown as a function of the ratio of $d_\mathrm{I}$ to a fixed $d$ for a 1D system without ISI using the Binomial distribution and the Poisson and Gaussian approximations.  It is observed from Fig.~\ref{fig:4} that the optimal detection interval can be very short compared to the transmission symbol interval. When $\Rx$ is much closer to $\Ix$ than to $\Tx$, i.e., $\dI < d$, a large $\Tr$ allows more molecules from $\Ix$ to be counted for the detection so $\Tr^\star$ should be much smaller than $\Tb$. Even when $\Rx$ is closer to $\Tx$ than to $\Ix$, i.e., $d<\dI$, but $\Ix$ is still close to $\Rx$, i.e., $\dI$ is small, $\Tr^\star$ should be much smaller than $\Tb$ to avoid molecules from $\Ix$. When $\Ix$ is farther from $\Rx$, i.e., $\dI$ increases, $\Tr^\star$  becomes larger. When $\Ix$ is very far from $\Rx$ as if it does not exist, we should have $\Tr=\Tb$ so that more molecules, which are only from $\Tx$, are counted for a more accurate detection. Moreover, when $\dI\approx d$, $\Tr=\Tb$ is  a good choice because molecules from $\Tx$ and $\Ix$ arrive at the receiver with equal probabilities and cannot be distinguished. Hence, taking all molecules into account can be helpful for the detection. Furthermore, we observe that the proposed algorithms accurately evaluates the global optimum since the optimal detection intervals, $\Tr^\star$, found by exhaustive search (shown by circle markers) match $\Tr^\star$ obtained by the algorithms. Only when $\dI<d$, the results for Poisson distribution are different from other results since $\pdi$ is large and the Poisson approximation is not accurate as explained in the discussion of Fig.~\ref{fig:2}.
	
	\begin{figure}
		\centering
		\includegraphics[width=0.45\textwidth]{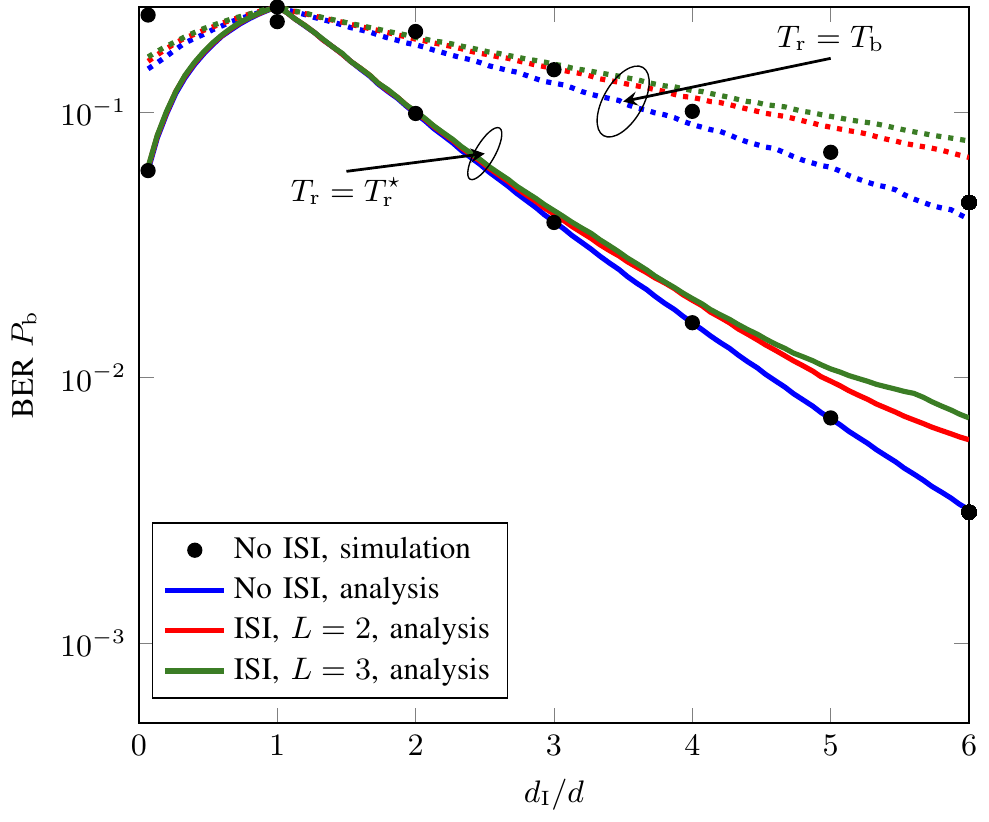}
		\caption{
			The BER of a 1D system as a function of the ratio of $d_\mathrm{I}$ to $d$ when $\Tr$ is optimized and when $\Tr=\Tb$. The systems with ISI use the optimal $\Tr$ designed for the non-ISI system.
		}
		\label{fig:5}
	\end{figure}
	
	In Fig.~\ref{fig:5}, we compare the BERs of  1D systems affected by external interference when $\Tr$ is optimal, i.e., $\Tr=\Tr^\star$, and when $\Tr=\Tb$. In particular, $\Tr^\star$ is designed  for the system without ISI and  the BERs of the systems  without ISI are presented. Moreover, the BERs of the systems with ISI, $L=2$ or $L=3$, using $\Tr^\star$ and $\Tr=\Tb$ are also presented.  From Fig.~\ref{fig:5}, we observe that when $\Tr=\Tb$, the BER, $\Pe$,  is much higher than the BER for $\Tr=\Tr^\star$.  The decrease in the BER for optimal $\Tr$ is more significant when the interference source is far away from the transmitter.  $\Pe=0.25$ when $\dI=d$ since information and interference molecules cannot be distinguished.   The simulation result confirms the analysis. Moreover, Fig.~\ref{fig:5} shows that the BERs of the systems with ISI are higher than that of the system without ISI, as expected. However, when the systems with ISI use   $\Tr^\star$ designed for the non-ISI system, their BERs are also reduced compared to the BER when using $\Tr=\Tb$. The reduction of BER due to the $\Tr^\star$ in the system with ISI is as significant as that in the system without ISI, for example, $10$ time reduction when $\dI/d=6$.

	\begin{figure}
		\centering
		\includegraphics[width=0.45\textwidth]{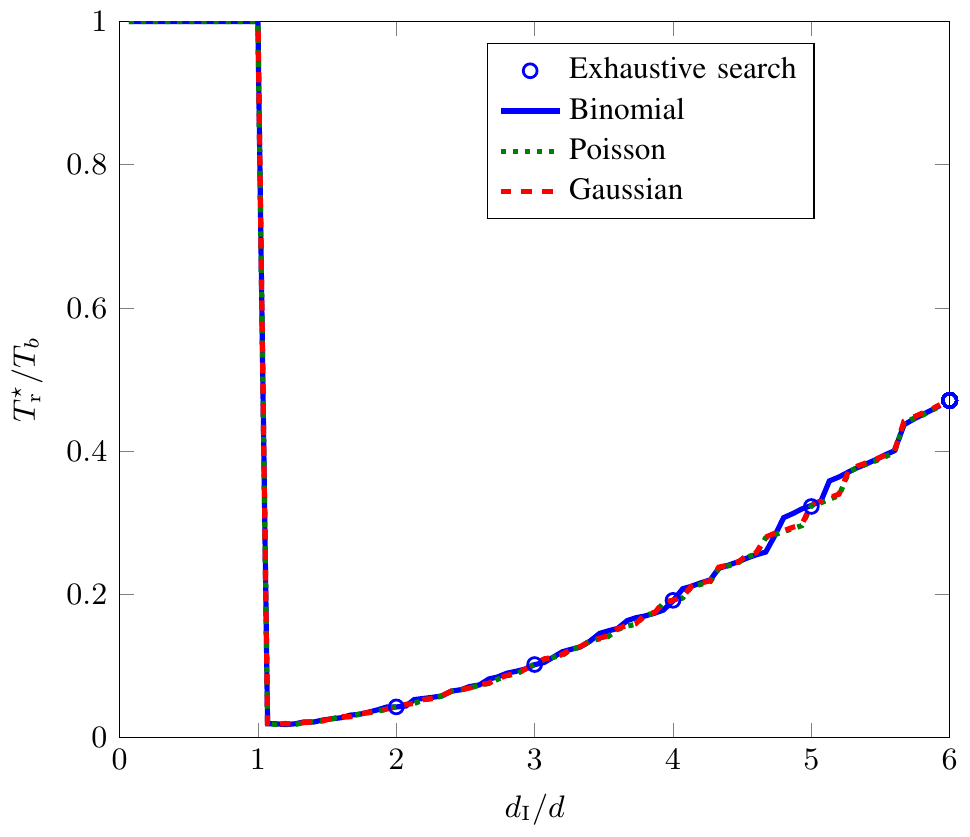}
		\caption{
			The ratio of the optimal detection  interval, $\Tr^\star$, to the transmission symbol interval, $\Tb$, as a function of the ratio of $d_\mathrm{I}$ to $d$ in a 3D system when using Binomial distribution, and Poisson and Gaussian approximations.
		}
		\label{fig:6}
	\end{figure}

	\begin{figure}
		\centering
		\includegraphics[width=0.45\textwidth]{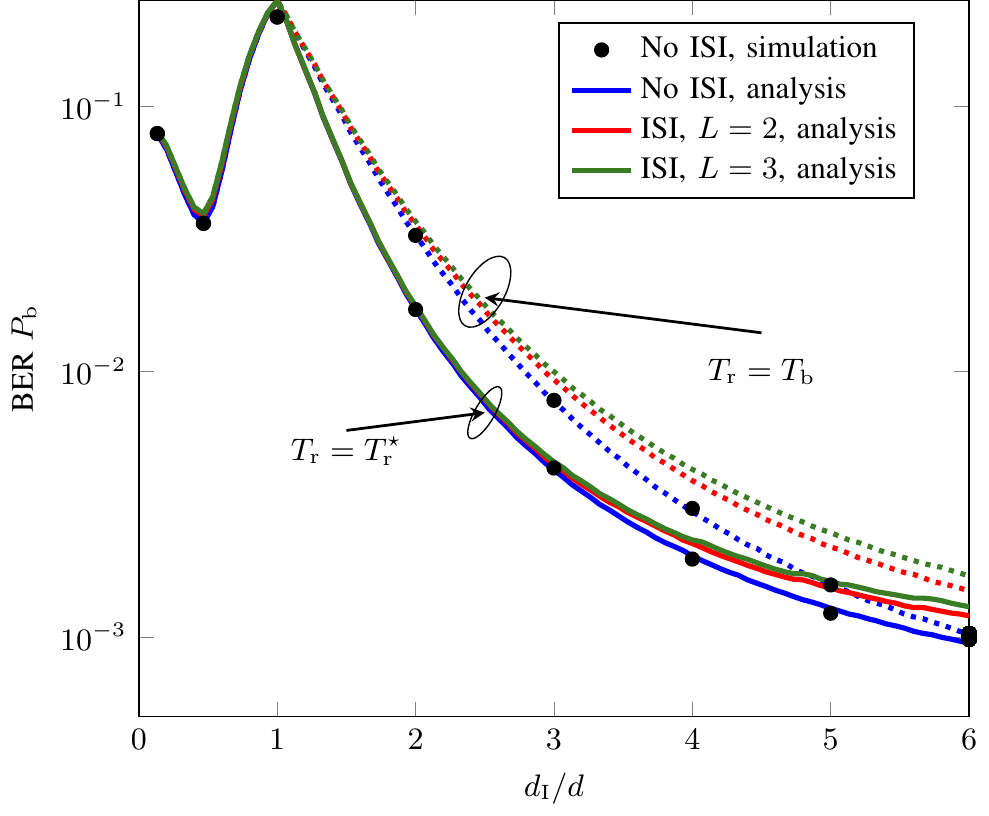}
		\caption{
			The BER of a 3D system as a function of the ratio of $d_\mathrm{I}$ to $d$ when $\Tr$ is optimized and when $\Tr=\Tb$, for ISI and no ISI
		}
		\label{fig:7}
	\end{figure}

	Fig.~\ref{fig:6} shows the ratio of the optimal detection  interval, $\Tr^\star$, to a fixed transmission symbol interval, $\Tb$,  as a function of the ratio of $d_\mathrm{I}$ to a fixed $d$ for the 3D  system without ISI  using Binomial distribution and the Poisson and Gaussian approximations.  We  observe from Fig.~\ref{fig:6} that $\Tr^{\star}$ can be much smaller than $\Tb$ when  $\Tx$ is closer to $\Rx$ than $\Ix$, i.e., $d<\dI$. The reason is similar to the 1D system, i.e., $\Tr^\star$ should be smaller than $\Tb$ so that fewer molecules from $\Ix$ are counted for the detection. However, when $\dI\leq d$, $\Tr^\star$ should be equal to $\Tb$, which is different from the 1D system. 
	The reason is that in a 3D system, $p_d$ and $\pdi$ are very small compared to the 1D system with the same parameters, which means even when $\Tr$ increases to infinity, all of the molecules cannot be received. Therefore, when $\dI\leq d$,  $\Tr=\Tb$ holds such that more molecules from the $\Tx$ can arrive for the detection, with the compromise of receiving more molecules from $\Ix$.   Moreover, the exhaustive search provides the same optimal $\Tr$ as $\Tr^\star$ given by the proposed algorithms. Poisson  and Gaussian approximations  give similar results to Binomial distribution since $p_d$ and $\pdi$ are small in 3D systems.

	Fig.~\ref{fig:7} shows the BER as a function of $\dI / d$ when $\Tr=\Tr^\star$ and $\Tr=\Tb$ for 3D systems without ISI and with ISI, i.e., $L=2$ or $L=3$. Note that $\Tr^\star$ is optimized for the system without ISI. In Fig.~\ref{fig:7}, we observe the improvement in the performance of systems with and without ISI in terms of BER by optimizing $\Tr$ compared to when $\Tr=\Tb$.  The improvement is significant when $\Ix$ is not too close or too far from $\Tx$, e.g., $\dI/d=3$.  If $\Ix$ is far from $\Tx$, molecules from $\Ix$ may not reach $\Rx$ and thus $\Tr^\star$ approaches $\Tb$, as shown in Fig.~\ref{fig:6}, and the improvement is not significant. If $\Ix$ is close to $\Tx$, more molecules from $\Ix$ are received and thus optimizing $\Tr$ is not helpful. Obviously, when $\dI\leq d$, $\Tr^\star=\Tb$ and thus there is no improvement in the BER.

	\begin{figure}
		\centering
		\includegraphics[width=0.45\textwidth]{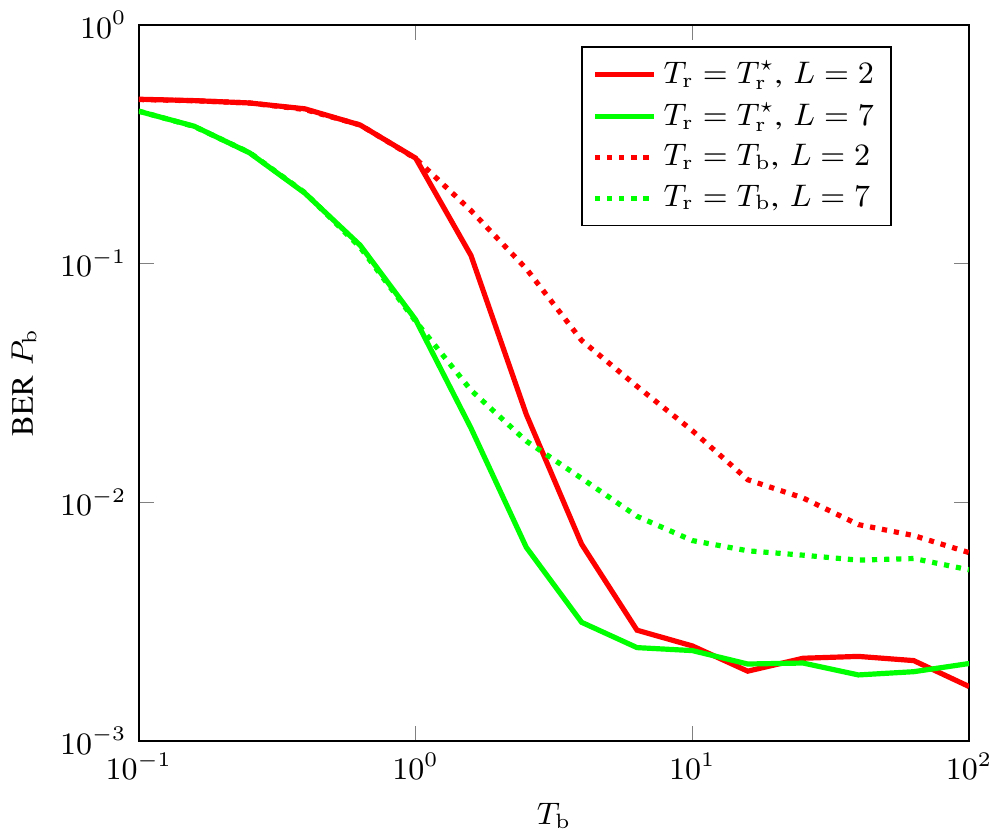}
		\caption{
			The BER of  3D systems with ISI using the detection that assumes $L=2$ or $L=7$ for $\Tr=\Tr^\star$ and $\Tr=\Tb$. 
		}
		\label{fig:8}
	\end{figure}

	Fig.~\ref{fig:8} shows the BER of 3D systems with ISI using the detection that assumes $L=2$ or $L=7$ as a function of $\Tb$ for $\Tr=\Tr^\star$ and $\Tr=\Tb$. $\Tr^\star$ is obtained by assuming no ISI in the system. The results are obtained by simulation. We consider 1000 sequences whose length is 100 symbols and ISI happens in the whole sequence. For a practical detection, the ML detections assume only ISI from one and six previous symbols, i.e., $L=2$ and $L=7$, respectively. In Fig.~\ref{fig:8}, we observe that when $\Tb$ is small, ISI dominates the inference from $\Ix$ and BER is high. Hence, in this case, optimizing $\Tr$ cannot improve the system performance. However, the BER of systems with ISI reduces significantly for $\Tr=\Tr^\star$ compared to $\Tr=\Tb$ when $\Tb$ is large, even though $\Tr^\star$ is optimized for the system without ISI. This is because ISI impact decreases when $\Tb$ increases. This confirms the benefit of the proposed optimal detection interval even in systems with ISI.

\begin{figure}
		\centering
		\includegraphics[width=0.45\textwidth]{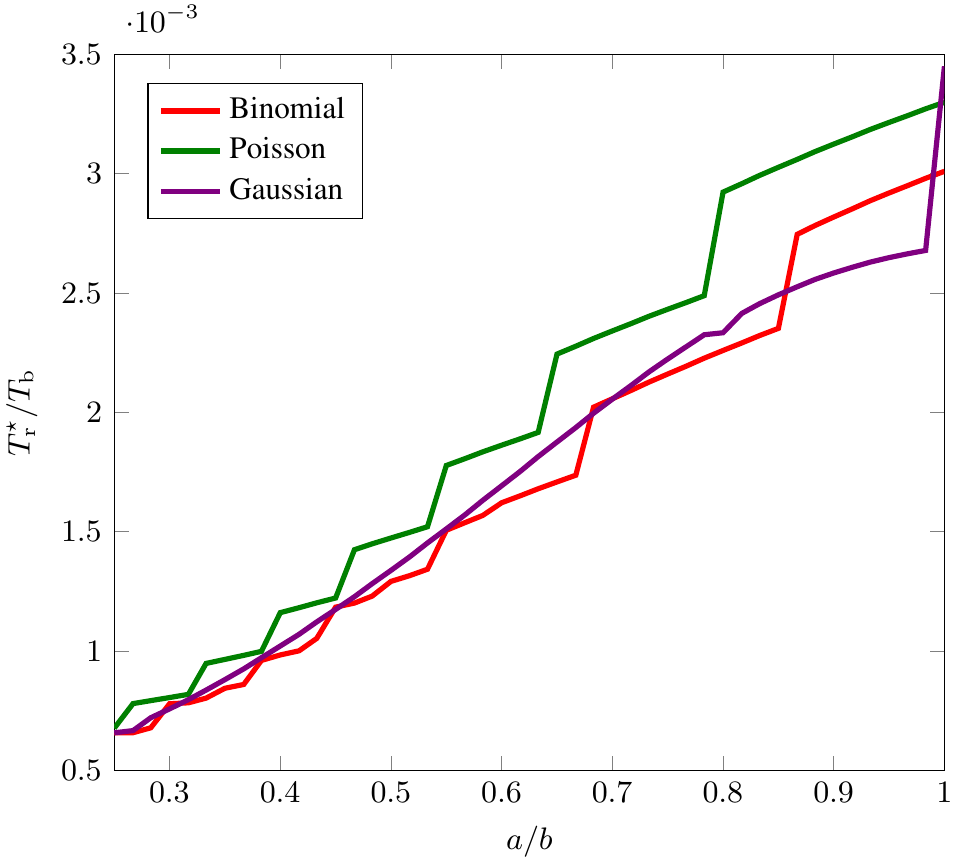}
		\caption{
			The ratio of the optimal detection  interval, $\Tr^\star$, to the transmission symbol interval, $\Tb$, as a function of $a/b$ in a 1D system with unknown-location interference. 
		}
		\label{fig:10}
	\end{figure}
\begin{figure}
		\centering
		\includegraphics[width=0.45\textwidth]{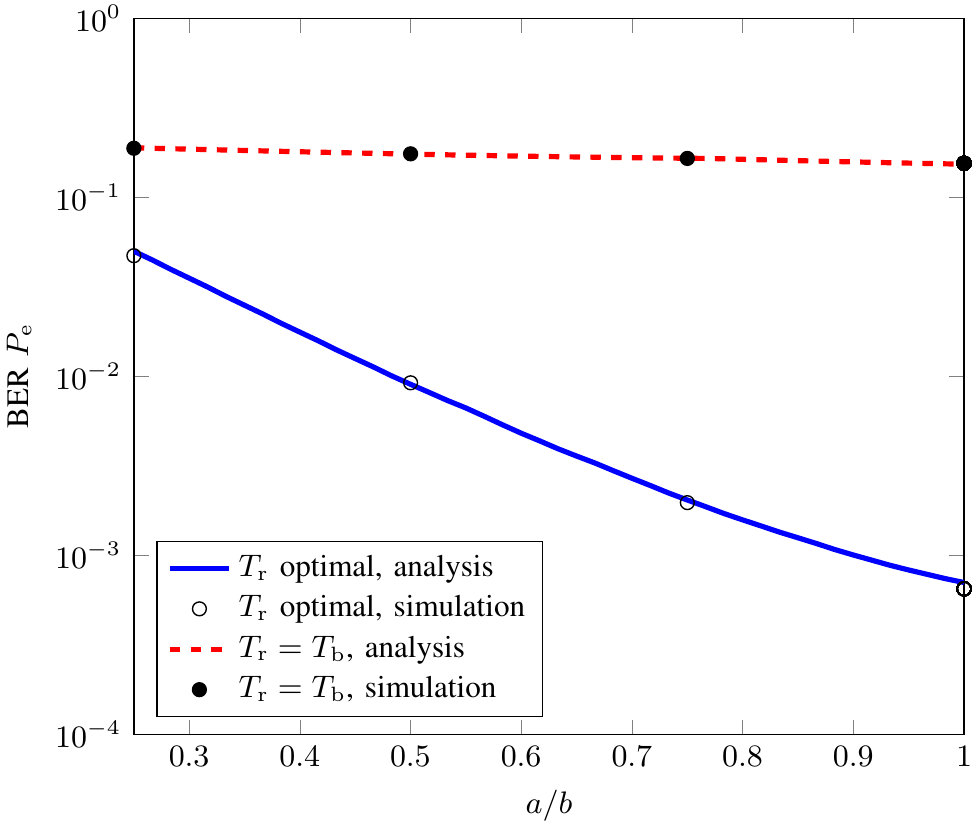}
		\caption{
			The BER as a function of $a/b$ when $\Tr$ is optimized and when $\Tr=\Tb$ in a 1D system with unknown-location interference.
		}
		\label{fig:11}
	\end{figure}

	Fig.~\ref{fig:10} presents the ratio of the optimal detection  interval, $\Tr^\star$, to the transmission symbol interval, $\Tb$,  as a function of $a/b$, when the  interference source is distributed uniformly between distances $a$ and $b$ from the receiver. Since the uncertain position of $\Ix$ reduces the system performance, we consider that $\Ix$ is far from the receiver compared to the transmitter so that the BER is not too high. We choose $a$ to vary from $\SI{3e-5}{\meter}$ to $\SI{12e-5}{\meter}$ and $b=\SI{12e-5}{\meter}$. As observed in Fig.~\ref{fig:10}, when $a$ and $b$ become close and the area where the interference source is located becomes further from the receiver, the ratio of $\Tr^\star$ to $\Tb$ increases.  The  BER of the system affected by interference at an unknown location  with optimal $\Tr$ is an improvement on the system with $\Tr=\Tb$, as shown in Fig.~\ref{fig:11}. As can be seen, the BER of the system with  optimal $\Tr$ is much lower than the BER  of the system with $\Tr=\Tb$.


	\section{Conclusion}\label{sec:7}
	
	In this paper, we investigated the optimal detection  interval at a receiver in a molecular communication system impaired by  external interference. In the 1D and 3D systems affected by external interference, our results showed that the optimal detection interval can be very small compared to the transmission interval. The BER is significantly reduced by optimizing the detection interval compared to when the detection interval is equal to the transmission interval. This also holds true for the system with ISI using the optimal detection interval of the system without ISI. Moreover, we have extended the 1D system model to  the case where the exact location of the interference source is unknown to the receiver. 
	The idea of optimizing the detection interval is simple but effective and thus practical for MC systems. Our results can also be extended to MC multi-access networks to improve the network performance and to mobile system where the transceivers and the interference are mobile, which can be considered for future work.

	\appendices
	
	\renewcommand{\thesectiondis}[2]{\Alph{section}:}
	\section{Proof of Lemma~\ref{lem:1}}\label{ap:0}
		
	When $\Tr$ changes within the interval $T_\mathrm{r}^{(l)}\leq\Tr\leq T_\mathrm{r}^{(l+1)}$, the values of $\PMF{Y|\Xt}{\left.y\right|x_\mathrm{T}=N_0}$ and $\PMF{Y|\Xt}{\left.y\right|x_\mathrm{T}=N_1}$ also change. However, the relation between $\PMF{Y|\Xt}{\left.y\right|x_\mathrm{T}=N_0}$ and $\PMF{Y|\Xt}{\left.y\right|x_\mathrm{T}=N_1}$, in terms of whether $\PMF{Y|\Xt}{\left.y\right|x_\mathrm{T}=N_0}>\PMF{Y|\Xt}{\left.y\right|x_\mathrm{T}=N_1}$ or $\PMF{Y|\Xt}{\left.y\right|x_\mathrm{T}=N_0}\leq\PMF{Y|\Xt}{\left.y\right|x_\mathrm{T}=N_1}$,  is preserved within the interval $T_\mathrm{r}^{(l)}\leq\Tr\leq T_\mathrm{r}^{(l+1)}$. Now, since $ \mathbb{Z}_0$ and $ \mathbb{Z}_1$ are the sets of discrete $Y$  obtained by comparing $\PMF{Y|\Xt}{\left.y\right|x_\mathrm{T}=N_0}$ and $\PMF{Y|\Xt}{\left.y\right|x_\mathrm{T}=N_1}$, the elements of $ \mathbb{Z}_0$ and $ \mathbb{Z}_1$ also do not change when $\Tr$  changes within the interval $T_\mathrm{r}^{(l)}\leq\Tr\leq T_\mathrm{r}^{(l+1)}$. 
		On the other hand, from \eqref{eq:6}, \eqref{eq:7}, \eqref{eq:27}, and \eqref{eq:28}, we can see that $p_d$ and $p_{d_\mathrm{I}}$ are smooth functions of $\Tr$. Hence, for $T_\mathrm{r}^{(l)}\leq\Tr\leq T_\mathrm{r}^{(l+1)}$, $l=1,2,\dots$, and $Y$ belonging to $ \mathbb{Z}_0$ or $ \mathbb{Z}_1$,   $\Pe$ in \eqref{eq:peb} is  a sum of smooth functions and therefore also a smooth function of $\Tr$ within this interval.   This can be proved strictly by taking the derivative of $\Pe$ with respect  to $\Tr$ when $Y$ belongs to the fixed sets, $ \mathbb{Z}_0$ and $ \mathbb{Z}_1$, and when $T_\mathrm{r}^{(l)}\leq\Tr\leq T_\mathrm{r}^{(l+1)}$ holds.  Note that, 	$\Pe$ is not smooth at the bounds of these intervals, i.e., $T_\mathrm{r}^{(l)}\leq\Tr\leq T_\mathrm{r}^{(l+1)}$.

	\section{Derivation of $P_{\mathrm{b},\textrm{Gaussian}}$ in \eqref{eq:peg}} \label {ap:1}
	
	To derive the BER from \eqref{eq:11}, we need to find $P_{\hat{X}_\mathrm{T}|\Xt}\left({\hat{x}_\mathrm{T}|x_\mathrm{T}}\right)$. Since  $\mathbb{Z}_0$ and $\mathbb{Z}_1$ are now continuous, we rewrite \eqref{eq:12} and \eqref{eq:13} as follows
	\begin{align} \label{eq:12a}
	&\PMF{\hat{X}_\mathrm{T}|\Xt}{\left.\hat{x}_\mathrm{T}=N_0\right|x_\mathrm{T}=N_1}\\\nonumber&\hspace{1cm}=\int_{y\in \mathbb{Z}_0} \PMF{Y|\Xt}{\left.y\right|x_\mathrm{T}=N_1}dy\nonumber\\ 
	&\hspace{1cm}\overset{(a)}{=}\frac{1}{2}\int_{y\in \mathbb{Z}_0} \PMF{Y|\Xt,X_\mathrm{I}}{\left.y\right|x_\mathrm{T}=N_1,x_\mathrm{I}=N_0}dy\nonumber\\ 
	&\hspace{1cm}\quad+\frac{1}{2}\int_{y\in \mathbb{Z}_0} \PMF{Y|\Xt,X_\mathrm{I}}{\left.y\right|x_\mathrm{T}=N_1,x_\mathrm{I}=N_1}dy,\nonumber
	\end{align}
	\begin{align} \label{eq:13a}
	&\PMF{\hat{X}_\mathrm{T}|\Xt}{\left.\hat{x}_\mathrm{T}=N_1\right|x_\mathrm{T}=N_0}\\\nonumber&\hspace{0.8cm}=\int_{y\in \mathbb{Z}_1} \PMF{Y|\Xt}{\left.y\right|x_\mathrm{T}=N_0}dy\nonumber\\ 
	&\hspace{0.8cm}\overset{(b)}{=}\frac{1}{2}\int_{y\in \mathbb{Z}_1} \PMF{Y|\Xt,X_\mathrm{I}}{\left.y\right|x_\mathrm{T}=N_0,x_\mathrm{I}=N_0}dy\nonumber\\ 
	&\hspace{0.8cm}\quad+\frac{1}{2}\int_{y\in \mathbb{Z}_1} \PMF{Y|\Xt,X_\mathrm{I}}{\left.y\right|x_\mathrm{T}=N_0,x_\mathrm{I}=N_1}dy,\nonumber
	\end{align}
	where  $(a)$ and $(b)$ follow from \eqref{eq:9} and \eqref{eq:1b}, respectively.

	Moreover, we have
	\begin{align}  \label{eq:22}
	&\int_{\gamma_{i}}^{\gamma_{i+1}}\PMF{Y|\Xt,X_\mathrm{I}}{\left.y\right|x_\mathrm{T},x_\mathrm{I}}dy=\\\nonumber
	&\hspace{0.5cm}\CDF{Y|\Xt,X_\mathrm{I}}{\left.\gamma_{i+1}\right|x_\mathrm{T},x_\mathrm{I}}-\CDF{Y|\Xt,X_\mathrm{I}}{\left.\gamma_{i}\right|x_\mathrm{T},x_\mathrm{I}},
	\end{align}
	where $\CDF{Y|\Xt,X_\mathrm{I}}{\left.\gamma\right|x_\mathrm{T},x_\mathrm{I}}$ is the cumulative distribution function (CDF)  of $Y$ given $\Xt$ and $\XI$. $\CDF{Y|\Xt,X_\mathrm{I}}{\left.\gamma\right|x_\mathrm{T},x_\mathrm{I}}$ is  given by
	\begin{align} \label{eq:23}
	&\CDF{Y|\Xt,X_\mathrm{I}}{\left.\gamma\right|x_\mathrm{T},x_\mathrm{I}}=\frac{1}{2}\\\nonumber
	&\hspace{1cm}\times\left(1+\erf\left(\frac{\gamma-\left(\Xt p_d+X_\mathrm{I}p_{d_\mathrm{I}}\right)}{\Xt p_d(1-p_d)+X_\mathrm{I}p_{d_\mathrm{I}}(1-p_{d_\mathrm{I}})}\right)\right).
	\end{align} 
	Therefore,
	$\mathbb{Z}_k=\underset{i}{\cup}\left[\gamma_{i} ,\gamma_{i+1}\right]$ can be written as
	\begin{align}  \label{eq:24}
	&\int_{y\in \mathbb{Z}_k} \PMF{Y|\Xt,X_\mathrm{I}}{\left.y\right|x_\mathrm{T},x_\mathrm{I}}dy\\\nonumber&\hspace{0.5cm}=\sum_{i}\left(\CDF{Y|\Xt,X_\mathrm{I}}{\left.\gamma_{i+1}\right|x_\mathrm{T},x_\mathrm{I}}-\CDF{Y|\Xt,X_\mathrm{I}}{\left.\gamma_{i}\right|x_\mathrm{T},x_\mathrm{I}}\right).
	\end{align}
	Inserting \eqref{eq:23} into \eqref{eq:24}, then \eqref{eq:24} into \eqref{eq:12a} and \eqref{eq:13a}, we obtain $P_{\hat{X}_\mathrm{T}|\Xt}\left({\hat{x}_\mathrm{T}|x_\mathrm{T}}\right)$. Then inserting $P_{\hat{X}_\mathrm{T}|\Xt}\left({\hat{x}_\mathrm{T}|x_\mathrm{T}}\right)$ and \eqref{eq:1a} into \eqref{eq:11},  we obtain the closed-form expression of the BER as in \eqref{eq:peg}.

	\section{Proof of global optimum $\Tr^\star$} \label{ap:5}
	
	To prove that $\Tr^\star$, obtained by Algorithm~\ref{al1}, for the Poisson distribution  is globally optimal, we need to prove that when  $\mathbb{Z}_0$ and  $\mathbb{Z}_1$ are fixed, $\Pe$ has only one local minimum. This is shown in the following. 
	
	\sloppy Since the sets $\mathbb{Z}_0$ and $\mathbb{Z}_1$ can be obtained by comparing $\PMF{Y|\Xt}{\left.y\right|x_\mathrm{T}=N_0}$ and  
	$\PMF{Y|\Xt}{\left.y\right|x_\mathrm{T}=N_1}$ for each $y$ in the interval $0 \leq y \leq 2N_1$, as shown by \eqref{eq:10}, $\mathbb{Z}_0$ and $\mathbb{Z}_1$ can be found by solving the following equation
	\begin{align} \label{eq:49}
	\PMF{Y|\Xt}{\left.y\right|x_\mathrm{T}=N_1}=\PMF{Y|\Xt}{\left.y\right|x_\mathrm{T}=N_0}.
	\end{align}
	If  equation \eqref{eq:49} has one and only one solution, denoted by $\gamma_\mathrm{th}$,   $\mathbb{Z}_0$ and  $\mathbb{Z}_1$ can be written as $\mathbb{Z}_0=\left\{0,\dots,\gamma_\mathrm{th}\right\}$ and  $\mathbb{Z}_1=\left\{\gamma_\mathrm{th}+1,\dots,2N_1\right\}$, respectively. Thus, we first prove that \eqref{eq:49} has one and only one solution, $\gamma_\mathrm{th}$. Then, we use $\gamma_\mathrm{th}$ to derive $\Pe$ and prove that  $\frac{\partial^2 \Pe}{\partial \Tr^2}>0$ with $\Tr$ satisfying $\frac{\partial \Pe}{\partial \Tr}=0$ when $\mathbb{Z}_0$ and  $\mathbb{Z}_1$ are fixed. Moreover, since $\Pe$ is continuous with respect to $\Tr$ when  $\mathbb{Z}_0$ and  $\mathbb{Z}_1$ are fixed, $\Pe$ has only one local minimal point.
	
	We set the left-hand side and the right-hand side of \eqref{eq:49} equal to a constant $m$, which is then presented by monotonic exponential functions. Thus, the solution of \eqref{eq:49} is the solution of the following set  of equations 
	\begin{align}\label{eq:050} 
	\begin{cases}
	&\left(N_1 p_d+N_0p_{d_\mathrm{I}}\right)^y e^{-\left(N_1 p_d+N_0p_{d_\mathrm{I}}\right)}+ \left(N_1 p_d+N_1p_{d_\mathrm{I}}\right)^y \\
		&\hspace{1cm}\times e^{-\left(N_1 p_d+N_1p_{d_\mathrm{I}}\right)}=m=u\left(N_1 p_d+N_1p_{d_\mathrm{I}}\right)^y  \\
	&\left(N_0 p_d+N_0p_{d_\mathrm{I}}\right)^y e^{-\left(N_0 p_d+N_0p_{d_\mathrm{I}}\right)}+ \left(N_0 p_d+N_1p_{d_\mathrm{I}}\right)^y \\
	&\hspace{1cm}\times e^{-\left(N_0 p_d+N_1p_{d_\mathrm{I}}\right)}=m=v\left(N_0 p_d+N_1p_{d_\mathrm{I}}\right)^y,
	\end{cases}
	\end{align}
	where $u,v$ are constants.
	Since each equation of the set in \eqref{eq:050} has only one solution, the solution of the set, i.e., the solution of \eqref{eq:49}, is unique.

	Now, from \eqref{eq:pep} and the unique $\gamma_\mathrm{th}$, we have
	\begin{align} \label{eq:50}
	\Pe&=\frac{1}{2}+\frac{1}{4}\left(\frac{\Gamma(\gamma_\mathrm{th}+1,N_1 p_d+N_0p_{d_\mathrm{I}})}{\gamma_\mathrm{th}!}\right.\nonumber \\&\quad+\frac{\Gamma(\gamma_\mathrm{th}+1,N_1 p_d+N_1p_{d_\mathrm{I}})}{\gamma_\mathrm{th}!}\nonumber \\ &\quad-\frac{\Gamma(\gamma_\mathrm{th}+1,N_0 p_d+N_0p_{d_\mathrm{I}})}{\gamma_\mathrm{th}!}\nonumber \\&\quad-\left.\frac{\Gamma(\gamma_\mathrm{th}+1,N_0 p_d+N_1p_{d_\mathrm{I}})}{\gamma_\mathrm{th}!}\right)
	\end{align}
	and
	\begin{align} \label{eq:51}
	&\frac{\partial \Pe}{\partial \Tr}=\frac{1}{4\gamma_\mathrm{th}!}\\\nonumber
	&\hfill\times\left(-\left(N_1 p_d+N_0p_{d_\mathrm{I}}\right)^{\gamma_\mathrm{th}}e^{-\left(N_1 p_d+N_0p_{d_\mathrm{I}}\right)}\left(N_1 p'_d+N_0 p'_{d_\mathrm{I}}\right)\right.\nonumber\\
	&-\left(N_1 p_d+N_1p_{d_\mathrm{I}}\right)^{\gamma_\mathrm{th}}e^{-\left(N_1 p_d+N_1p_{d_\mathrm{I}}\right)}\left(N_1 p'_d+N_1 p'_{d_\mathrm{I}}\right)\nonumber\\
	&+\left(N_0 p_d+N_0p_{d_\mathrm{I}}\right)^{\gamma_\mathrm{th}}e^{-\left(N_0 p_d+N_0p_{d_\mathrm{I}}\right)}\left(N_0 p'_d+N_0 p'_{d_\mathrm{I}}\right)\nonumber \\
	&\left.+\left(N_0 p_d+N_1p_{d_\mathrm{I}}\right)^{\gamma_\mathrm{th}}e^{-\left(N_0 p_d+N_1p_{d_\mathrm{I}}\right)}\left(N_0 p'_d+N_1 p'_{d_\mathrm{I}}\right)\right).\nonumber
	\end{align}
	When $\frac{\partial \Pe}{\partial \Tr}=0$, we have
	\begin{align} \label{eq:52}
	&\left(N_1 p_d+N_0p_{d_\mathrm{I}}\right)^{\gamma_\mathrm{th}}e^{-\left(N_1 p_d+N_0p_{d_\mathrm{I}}\right)}\left(N_1 p'_d+N_0 p'_{d_\mathrm{I}}\right)=\nonumber\\
	&-\left(N_1 p_d+N_1p_{d_\mathrm{I}}\right)^{\gamma_\mathrm{th}}e^{-\left(N_1 p_d+N_1p_{d_\mathrm{I}}\right)}\left(N_1 p'_d+N_1 p'_{d_\mathrm{I}}\right)\nonumber\\
	&+\left(N_0 p_d+N_0p_{d_\mathrm{I}}\right)^{\gamma_\mathrm{th}}e^{-\left(N_0 p_d+N_0p_{d_\mathrm{I}}\right)}\left(N_0 p'_d+N_0 p'_{d_\mathrm{I}}\right)\nonumber \\
	&+\left(N_0 p_d+N_1p_{d_\mathrm{I}}\right)^{\gamma_\mathrm{th}}e^{-\left(N_0 p_d+N_1p_{d_\mathrm{I}}\right)}\left(N_0 p'_d+N_1 p'_{d_\mathrm{I}}\right).
	\end{align}
	From \eqref{eq:51}, we can derive $\frac{\partial^2 \Pe}{\partial \Tr^2}$. Then, substituting \eqref{eq:52} into $\frac{\partial^2 \Pe}{\partial \Tr^2}$, we see that
	\begin{align}
	\frac{\partial^2 \Pe}{\partial \Tr^2}> 0.
	\end{align}
	Hence, the stationary point of $\Pe$ is a minimum. On the other hand, since $\Pe$ is continuous when  $\mathbb{Z}_0$ and  $\mathbb{Z}_1$ are fixed, $\Pe$ has only one minimal point and thus the optimal point given by Algorithm~1 is  global optimal.

	\bibliographystyle{IEEEtran}
	\bibliography{IEEEabrv,MolecularBib}

\end{document}